\documentclass{jpp}

\usepackage{graphicx}

\usepackage{amsmath}
\usepackage{amssymb}
\usepackage{epstopdf}
\usepackage[lofdepth,lotdepth]{subfig}
\usepackage{tabularx,multirow}
\usepackage[makeroom]{cancel}
\usepackage{color}
\usepackage[colorlinks=true,linkcolor=blue,citecolor=blue]{hyperref}

\ifCUPmtlplainloaded \else
  \checkfont{eurm10}
  \iffontfound
    \IfFileExists{upmath.sty}
      {\typeout{^^JFound AMS Euler Roman fonts on the system,
                   using the 'upmath' package.^^J}%
       \usepackage{upmath}}
      {\typeout{^^JFound AMS Euler Roman fonts on the system, but you
                   dont seem to have the}%
       \typeout{'upmath' package installed. JPP.cls can take advantage
                 of these fonts, if you use 'upmath' package.^^J}%
      }
  \else
  \fi
\fi

\ifCUPmtlplainloaded \else
  \checkfont{msam10}
  \iffontfound
    \IfFileExists{amssymb.sty}
      {\typeout{^^JFound AMS Symbol fonts on the system, using the
                'amssymb' package.^^J}%
       \usepackage{amssymb}%

      }{}
  \fi
\fi

\ifCUPmtlplainloaded \else
  \IfFileExists{amsbsy.sty}
    {\typeout{^^JFound the 'amsbsy' package on the system, using it.^^J}%
     \usepackage{amsbsy}}
    {\providecommand\boldsymbol[1]{\mbox{\boldmath $##1$}}}
\fi

\newcommand{\figref}[1]{figure~\ref{#1}}

\newcommand{\Figref}[1]{Figure~\ref{#1}}

\newcommand{\secref}[1]{\S\ref{#1}}

\newcommand{\exref}[1]{(\ref{#1})}

\newcommand{\beq}{\begin{equation}}
\newcommand{\eeq}{\end{equation}}
\newcommand{\bea}{\begin{eqnarray}}
\newcommand{\eea}{\end{eqnarray}}

\newcommand{\mbf}[1]{\boldsymbol{#1}}

\newcommand{\vc}[1]{\mbf{#1}}

\newcommand{\dd}{\partial}

\newcommand{\mfp}{\lambda_{\rm mfp}}

\newcommand{\vth}{v_{\rm th}}

\newcommand{\vthi}{v_{{\rm th}i}}
\newcommand{\vthe}{v_{{\rm th}e}}

\newcommand{\kpl}{k_{\parallel}}
\newcommand{\kpd}{k_{\perp}}
\newcommand{\vpl}{v_{\parallel}}
\newcommand{\vpd}{v_{\perp}}

\newcommand{\Sw}{S_{\rm w}}

\title[Thermal conduction in whistler-unstable plasma]{Self-inhibiting thermal conduction in high-beta, whistler-unstable plasma}
\author[S.\ Komarov, A.\ Schekochihin, E.\ Churazov and A.\ Spitkovsky]{
S.~Komarov$^{1}$\thanks{E-mail: komarov@mpa-garching.mpg.de}, 
A.~Schekochihin$^{2,3}$, 
E.~Churazov$^{4}$ 
 and A.~Spitkovsky$^{5}$}
\affiliation{
$^{1}$Space Research Institute (IKI), Profsouznaya 84/32, Moscow 117997, Russia\\
$^{2}$The Rudolf Peierls Centre for Theoretical Physics, University of Oxford, 1 Keble Road,\\
Oxford OX1 3NP, UK
\\[\affilskip]
$^{3}$Merton College, Oxford OX1 4JD, UK
\\[\affilskip]
$^{4}$Max Planck Institute for Astrophysics, Karl-Schwarzschild-Strasse 1, 
85741 Garching, Germany
\\[\affilskip]
$^{5}$Department of Astrophysical Sciences, Princeton University, Peyton Hall, Princeton,\\
NJ 08544, USA} 

\begin{document}
\maketitle
\label{firstpage}

\begin{abstract}
A heat flux in a high-$\beta$ plasma with low collisionality triggers the whistler instability. 
Quasilinear theory predicts saturation of the instability in a  marginal state characterized by a heat flux that is fully controlled by electron scattering off magnetic perturbations. This 
marginal heat flux does not depend on the temperature gradient and scales as $1/\beta$. 
We confirm this theoretical prediction by performing numerical particle-in-cell simulations of 
the instability. We further calculate the saturation level of magnetic perturbations and the electron scattering rate as functions of $\beta$ and the temperature gradient to identify the saturation mechanism as quasilinear. 
Suppression of the heat flux is caused by oblique whistlers with magnetic-energy density distributed over a wide range of propagation angles. This result can be applied to high-$\beta$ 
astrophysical plasmas, such as the intracluster medium, where thermal conduction at sharp 
temperature gradients along magnetic-field lines can be significantly suppressed. We provide 
a convenient expression for the amount of suppression of the heat flux relative to the classical 
Spitzer value as a function of the temperature gradient and $\beta$. For a turbulent plasma, 
the additional independent suppression by the mirror instability is capable of producing large total suppression factors (several tens in galaxy clusters) in regions with strong temperature gradients.
\end{abstract}


\section{Introduction}
\label{sec:intro}
Thermal conduction in a hot magnetized turbulent astrophysical plasma has been an 
active research topic since its potential role in the thermodynamics of galaxy 
clusters was appreciated (the so-called cooling flow problem; see, e.g., \citealt{
Rusz2002,VoigtFabian2004,Zakamska2003,Dennis2005}). It is often assumed to be 
significantly reduced relative to the unmagnetized Spitzer 
conductivity by magnetic fields, an assumption mainly based on observations of various temperature substructure 
in the intracluster medium (ICM): fluctuations on scales of the order of 100 kpc 
\citep{Mark2003,Wang2016}, sharp gradients at contact discontinuities (the so-called cold fronts; e.g., \citealt{
Mark2000,EttoriFabian2000,Vikhl2001,MarkVikhl2007,Ichi2015}), and filamentary structures (e.g., in the Coma cluster; see \citealt{Sanders2013}). However, the exact physics 
of such suppression remain to be understood. To some extent, the persistence of 
temperature fluctuations could be explained by the turbulent topology of magnetic-field lines that favors perpendicular orientation of temperature gradients and 
field lines \citep{Komarov2014}, while cold fronts likely survive for long times 
due to field-line draping, which has similar effect by stretching field lines 
along a cold front interface \citep{Lyutikov2006,Asai2007,Dursi2008}. 

It is in general more problematic to suppress parallel thermal conduction along magnetic-field 
lines. In order to inhibit parallel electron transport, small-scale magnetic 
fluctuations that presumably exist in the ICM due to various kinetic instabilities 
should be either in the form of transverse perturbations on electron Larmor scales 
\citep[e.g.,][]{LE1992}, or in the form of magnetic mirrors (i.e., longitudinal waves) 
on larger scales \citep[e.g.,][]{CC1998}. 
The latter may be provided on ion Larmor scales by the mirror instability, driven by anisotropy in the plasma temperature that is biased perpendicularly with respect to the local magnetic-field direction 
(\citealt{Parker,Hasegawa}; see also \citealt{Kunz2014} and \citealt{Rincon2015}
for the saturation mechanism).
Suppression factors estimated for this case are rather 
modest, of the order of 1/3-1/5 of the Spitzer value \citep{Komarov2016}. 

The transverse whistler instability seems to be the most promising 
candidate for scattering electrons at the scale of their Larmor radii. It has long been 
known that a heat flux in a weakly collisional magnetized plasma causes whistler instability 
under certain conditions and thus can, possibly, inhibit itself (\citealt{LE1992}; see also 
\citealt{RL1978} for the unmagnetized case). This problem presents significant 
theoretical interest, even outside of the context of galaxy clusters. \cite{LE1992} 
first described the linear heat-flux-induced whistler instability and estimated the 
suppression of thermal conduction by assuming that saturation of the instability is 
controlled by nonlinear mode coupling. In their work, they employed the simple isotropic 
Krook operator in order to describe electron scattering off whistler perturbations. 
\cite{PE1998} (hereafter PE98) questioned the  validity of this assumption and 
demonstrated that in the framework of quasilinear theory (QLT), the marginal electron 
distribution function in fact generates oblique whistlers able to scatter heat-carrying 
electrons efficiently. Both \cite{LE1992} and PE98 stressed the fact that 
strictly parallel whistler modes do not interact with heat-carrying 
electrons intensively, because the Doppler-shifted circular rotation of the E-vector of these 
modes in the frame moving with the electrons along the mean magnetic field is opposite 
to the gyration of the electrons. The elliptical polarization of oblique modes, on the 
other hand, alleviates this problem and enables resonant interaction with the heat-carrying 
particles. The resulting heat flux in PE98 turns out to be independent of the 
temperature gradient and scales as the inverse electron plasma beta, $\beta_e^{-1}$. 

In this work, we study the heat-flux-induced whistler instability with the aid of
particle-in-cell numerical simulations. By performing runs with different values of $\beta_e$ and temperature 
gradients, we arrive at qualitatively the same conclusion as PE98: the saturated 
whistler modes are oblique and, therefore, successfully inhibit the electron heat flux, 
restricting it to the $\beta_e^{-1}$ scaling regardless of the magnitude of the temperature 
gradient. We also show that the saturated magnetic-field energy, as well as the pitch-angle 
scattering rate follow the 
same functional form as predicted by QLT.

During the final stage of preparation of this paper, \cite{RC2017} published 
a very similar numerical result (see also \citealt{RC2016} for their 
previous work on this subject). Our work can be considered as an independent confirmation 
of their main result, namely, the fact that the heat flux controlled by the instability scales 
as $\beta_e^{-1}$. However, we propose a rather different physical approach to the interpretation of this result, based on 
quasilinear saturation near marginal stability. In addition, we discuss some of the aspects of the 
instability in more detail, e.g., the structure of the electron distribution function 
in the marginal state and the scaling of the pitch-angle scattering rate and  
saturation level of magnetic perturbations with $\beta_e$ and the temperature 
gradient. We also provide a convenient expression for the suppression factor of 
the heat flux applicable to clusters of galaxies. This model could be easily
 incorporated into hydro- and magnetohydrodynamic numerical simulations. 

The rest of this paper is organized as follows. In \secref{sec:phys}, we present a qualitative explanation 
of the physics behind the heat-flux suppression by whistler turbulence based on the 
marginality criterion. We then turn to numerical results (\secref{sec:num}) to support this 
model. We proceed with discussion of the relevance of our results to galaxy clusters and 
the limitations of our model in \secref{sec:disc}. We summarize our findings in \secref{sec:concl}. 

\section{Theoretical considerations}
\label{sec:phys}

\subsection{General remarks}
\label{sec:general}
Let us assume that due to a certain anisotropy of the electron distribution function in a weakly collisional magnetized plasma, it becomes unstable 
and triggers electromagnetic modes propagating in some direction with phase 
velocities $\omega/k$, where $\omega$ is the wave frequency and $k$ is the wavenumber. 
We assume also that the electrons are fast compared to the wave (as tends to be the case for for low-frequency modes 
in a hot plasma), so the electron Landau resonance is ineffective and wave-particle interactions 
mostly happen via gyroresonances $k_{\parallel} v_{\parallel}=\pm \Omega_e$, where $\kpl$ is the parallel (to the mean magnetic field) wavenumber, $v_{\parallel}$ the parallel electron velocity, and $\Omega_e$ the electron Larmor frequency. Let the unstable modes have random phases and a 
sufficiently broad spectrum, viz., $\Delta k/k \sim 1$. This allows electrons within a wide range 
of parallel velocities to resonate with different uncorrelated 
wave modes. 
For an electromagnetic wave, the perpendicular electric field $\delta E_{\perp}'$ in the reference frame moving at the parallel phase velocity of the wave is zero: 
\beq
\label{eq:Eperp}
\delta E_{\perp}' = \delta E_{\perp} - \omega/(\kpl c) \delta B_{\perp} = \delta E_{\perp} - [\omega/(\kpl c)] (c/\omega) \kpl \delta E_{\perp} = 0,
\eeq 
where  $\delta B_{\perp}$ and $\delta E_{\perp}$ are the perpendicular 
magnetic and electric fields of the wave in the lab frame, and $c$ is the speed of light. 
The parallel electric field (unaffected by the change of the reference frame) can be safely assumed unimportant because it only interacts with electrons 
via Landau resonance, but because the wave is slow compared to the electron thermal speed, 
such interaction is weak, and there is no secular change in a particle's energy\footnote{More precisely, it can be shown by a calculation similar to \exref{eq:Eperp} that the vector potential $\vc{A}$ 
of the wave is zero in the frame of reference moving with the parallel phase velocity, while the electrostatic potential $\phi$ is typically very small. For whistlers, $\phi\gtrsim(v/c)A$ only close to the resonant cone whose opening angle approaches $\pi/2$ at low frequencies far below the electron plasma frequency. Such quasielectrostatic modes should not be excited by electron distribution functions with 
a negative velocity-magnitude derivative.}.
Resonant particles, which are the ones that mostly contributed to the initial anisotropy (the instability drains free energy from the anisotropy by resonant wave-particle interactions), are 
scattered elastically by magnetic perturbations in the moving frame. Eventually, this leads to isotropization of their distribution function in the `wave frame' and quenching 
of the instability. Thus,  an excess of particles at parallel momenta in the direction of the wave propagation larger than of the order of $m_e \omega / \kpl$ in the lab frame is not allowed. 

Let us assume for illustrative purposes that the electrons have a non-zero mean momentum in the direction of the wave, causing an asymmetry in the electron distribution function. 
We may define the anisotropy of such distribution simply as $\epsilon = \langle p_{\parallel} \rangle /
 p_{\rm th}$, where  $\langle p_{\parallel} \rangle$ is the mean parallel momentum and $p_{\rm th} = (2m_e
  T)^{1/2}$ the electron thermal momentum (we use energy units of temperature everywhere). Then, the instability limits such anisotropy by  $\epsilon_{\rm max}\sim \omega/\kpl \vth$, where $\vth$ is 
the electron thermal velocity. A parallel heat flux is, in fact, characterized by a similar perturbation of the distribution function\footnote{With the important exception that a plasma produces a flow of colder particles opposite to the direction of the heat flux to cancel the electric current and make $\langle p_{\parallel} \rangle = 0$. The part of the distribution function associated with such backflow, however, does not play a key role in the instability, as will be shown in \secref{sec:margheatflux}. We therefore use a simplified model with $\langle p_{\parallel} \rangle \neq 0$ in this section for illustrative reasons. } 
at parallel velocities 
 $v_{\parallel} \sim \pm \vth$. We can roughly estimate the heat flux as
\beq
q_{\parallel} \sim m_e n \vth^2 \langle v_{\parallel} \rangle \sim \epsilon m_e n \vth^3  
\eeq
where $n$ is 
the electron density. For the heat-carrying particles, the resonant 
scale is simply the electron Larmor radius $\rho_e=\vth/\Omega_e$, which follows from the gyroresonance condition $\kpl \vpl = \pm \Omega_e$.  The frequency of whistler waves at 
this scale is 
\beq
\omega\sim (\Omega_e/\omega_{\rm p}^2) k^2 c^2
\sim (k \rho_e)^2 \Omega_e / \beta_e   \sim \Omega_e / \beta_e.
\eeq 
Thus, the whistler phase 
velocity is $\vth/\beta_e$, and, immediately, if whistler turbulence 
saturates by electron pitch-angle scattering, it limits the maximum anisotropy to $\sim 1/\beta_e$. 
Equivalently, the marginal heat flux should be
\beq
\label{eq:genheatflux}
q_{\parallel} \sim \beta_e^{-1} m_e n \vth^3,
\eeq
provided that such flux turns out to be smaller than the initial heat flux with no instability. 
Already from these simplified arguments, one gets a heat flux that is fully controlled by the 
plasma beta and is independent of the imposed temperature gradient. This is exactly 
the conclusion made by PE98 via a more rigorous quasilinear derivation. 

\subsection{Whistler instability}

It is most convenient to establish the connection between the electron distribution 
function and the growth rate with the help of basic semi-classical concepts 
\citep{Melrose1980}. This method is physically equivalent to the usual derivation 
based on the Landau-Laplace procedure. We prefer this derivation because it provides a quick shortcut to the expression for the growth rate in a form that allows one to study marginal 
distribution functions without knowing the complicated details of the dispersion relation in 
the general case of oblique wave propagation. We also believe it is clearer for a reader 
less familiar with plasma kinetics, because it does not introduce from the start the dispersion relation, 
which is often hard to interpret physically.  
 
Let an electron with momentum $\vc{p}$ gyrating in a magnetic field emit a photon 
with momentum $\hbar \vc{k}$. The change in the electron's parallel momentum 
is 
\beq
\label{eq:parmom}
\Delta p_{\parallel} = -\hbar k_{\parallel}.
\eeq 
Conservation of energy implies
\beq
\label{eq:encons}
\Delta \frac{p_{\parallel}^2}{2m_e} + \Delta \frac{p_{\perp}^2}{2m_e} + \hbar \omega=0.
\eeq
The perpendicular kinetic energy of an electron in a magnetic field is quantized 
as $E_{\perp}=j \hbar \Omega_e$, where $j$ is a non-negative integer (we can ignore 
the electron's spin and ground-state energy here as we are interested in the classical limit 
$j \gg 1$). Then the change in the perpendicular momentum of the electron is 
\beq
\label{eq:perpmom}
\Delta p_{\perp}=-s\hbar\Omega_e/v_{\perp},
\eeq
where $s$ is an integer. From \exref{eq:encons}, using \exref{eq:parmom}, we get
\beq
\label{eq:rescond}
\omega - k_{\parallel} v_{\parallel}=s\Omega_e.
\eeq
This is just the normal resonance condition, which is in fact the statement of 
energy conservation. 

The number of emitted photons with wave vector $\vc{k}$ per unit time between 
two electron states with momenta $\vc{p}$ and $\vc{p}-\hbar \vc{k}$ is set by the difference 
between the rates of stimulated emission and stimulated absorption. The 
former should be proportional to the electron distribution function at 
the higher momentum $\vc{p}$, $f(\vc{p})$, and the latter to $f(\vc{p}-\hbar\vc{k})$. 
Assuming that the emitted/absorbed momentum is small and using \exref{eq:parmom} 
and \exref{eq:perpmom}, we get 
\beq
\Delta f(\vc{p}, \vc{k}) = f(\vc{p}) - f(\vc{p}-\hbar \vc{k}) = -\Delta p_{\perp} \frac{\partial f}{\dd p_{\perp}} 
- \Delta p_{\parallel} \frac{\partial f}{\partial p_{\parallel}} 
= \hbar \left ( \frac{s\Omega_e}{v_{\perp}} \frac{\partial}{\partial p_{\perp}} + k_{\parallel} \frac{\partial}{\partial p_{\parallel}} \right ) f(\vc{p}).
\label{eq:deltaf}
\eeq
Now we can obtain the rate of energy transfer from the electrons to the wave 
by integrating over electron momenta:
\beq
\frac{d \mathcal{E}(\vc{k})}{dt}= \int d^3 \vc{p} \ w(\vc{p},\vc{k}) \Delta f(\vc{p}, \vc{k}) \mathcal{E}(\vc{k}),
\eeq
where $\mathcal{E}(\vc{k})$ is the density of energy contained in the wave, $w(\vc{p},\vc{k})= 
\mathcal{W}(\vc{p},\vc{k})\delta(p_{\parallel} - p_{\parallel {\rm r}})$ is the probability 
of stimulated emission/absorption of a photon with wave vector $\vc{k}$ by an electron with 
momentum $\vc{p}$ per unit of time, and $p_{\parallel \rm r}= m_e (\omega - s \Omega_e) / 
k_{\parallel}$ is the resonant parallel momentum. 
The non-negative function $\mathcal{W}(\vc{p},\vc{k})$ contains all the information 
about the dispersion relation of the particular emitted mode. 
The wave energy growth rate $\gamma_s(\vc{k})$ is then
\beq
\gamma_s(\vc{k}) = \int d^3 \vc{p} \ w(\vc{p},\vc{k}) \Delta f(\vc{p},\vc{k}).
\eeq
Here and in what follows, the subscript $s$ indicates that the growth rate $\gamma_s$ is given for a single-$s$ resonance \exref{eq:rescond}, while the total growth rate is an infinite sum over $s$.    
Using \exref{eq:deltaf}, we finally arrive at a general expression for the growth 
rate of an arbitrary electromagnetic mode in the form
\beq
\gamma_s(\vc{k}) = \int  d^3 \vc{p} \ \delta(p_{\parallel} - p_{\parallel {\rm r}}) \mathcal{W}(\vc{p},\vc{k})
	 \hbar \left ( \frac{s\Omega_e}{v_{\perp}} \frac{\dd}{\dd p_{\perp}} + k_{\parallel} \frac{\dd}{\dd p_{\parallel}} \right ) f(\vc{p}).
\label{eq:gammak}
\eeq
Expression~\exref{eq:gammak} is 
convenient in the sense that it is valid for waves propagating at arbitrary angles, 
and it allows one to link the sign of the growth rate to properties of the distribution 
function without knowing the complicated dispersion relation for the general case of 
oblique propagation. 
The sign of $\gamma_s(\vc{k})$ is determined by function
\beq
\label{eq:Gamma}
\Gamma_s(p_{\perp},\vc{k}) = \left ( \frac{s\Omega_e}{v_{\perp}} \frac{\dd}{\dd p_{\perp}} + k_{\parallel} \frac{\dd}{\dd p_{\parallel}} \right ) f(\vc{p}) \vert_{p_{\parallel}=p_{\parallel {\rm r}}}.
\eeq  
Switching to velocity derivatives and using the resonance condition \exref{eq:rescond}, we get
\beq
\Gamma_s(v_{\perp},\vc{k}) \propto \frac{\kpl}{|\kpl|} \left [- \left ( v_{\parallel} - \frac{\omega}{k_{\parallel}} \right ) \frac{\dd}{\dd v_{\perp}} + v_{\perp} \frac{\dd}{\dd v_{\parallel}}\right ] 
 f(\vc{v}) \vert_{v_{\parallel}=v_{\parallel {\rm r}}}.
\label{eq:gammapart}
\eeq
Note that the sum of the partial derivatives 
in \exref{eq:gammapart} is a derivative taken along semicircles 
$(v_{\parallel}-\omega/k_{\parallel})^2+v_{\perp}^2={\rm const}$. This represents the fact 
that electron energy is conserved in the frame moving with the parallel phase velocity of the wave 
(because the perpendicular electric field is zero there). 
Equation~\exref{eq:gammapart} can be cast in a compact form:
\beq
\label{eq:gammacircle}
\Gamma_s(v_{\perp}, \vc{k}) \propto \frac{\kpl}{|\kpl|} \hat{\vc{l}} \cdot \frac{\dd f}{\dd \vc{v}} \bigg \vert_{v_{\parallel}=v_{\parallel {\rm r}}},
\eeq
where $\hat{\vc{l}}=(\sin \phi, - \cos \phi)$ is a unit vector pointing clockwise along 
the contours of constant energy in the wave frame, and $\phi$ is the polar angle in coordinates 
$(v_{\parallel}-\omega/k_{\parallel}, v_{\perp})$. Let us choose $\kpl>0$ 
without loss of generality. We see that 
instability occurs when the distribution function near the resonant parallel momenta 
increases in the clockwise direction along the equi-energy contours in the wave frame. 
For the resonance at $v_{\parallel} \approx - \Omega_e/k_{\parallel}$, a parallel momentum 
deficiency (or, equivalently, a surplus of particles with high $v_{\perp}$) is needed for the instability 
to occur, while at $v_{\parallel}\approx \Omega_e/k_{\parallel}$ one needs an excess of parallel momentum (see \figref{fig:margin} for an illustration). The case of $\kpl<0$ is analogous and described 
by the oriented contours of constant energy in \figref{fig:margin} mirror-reflected with respect to the $y$-axis, i.e., the direction of positive wave 
growth changes to counterclockwise.

\subsection{Marginal heat flux}
\label{sec:margheatflux}

Provided that the spectrum of excited modes is sufficiently broad ($\Delta k_{\parallel}/k_{\parallel}\sim 1$), 
particles in a wide range of parallel velocities are scattered by magnetic perturbations, and their 
isotropization in the wave frame leads to marginal stability.

\begin{figure*}
\centering
\includegraphics[width=130mm]{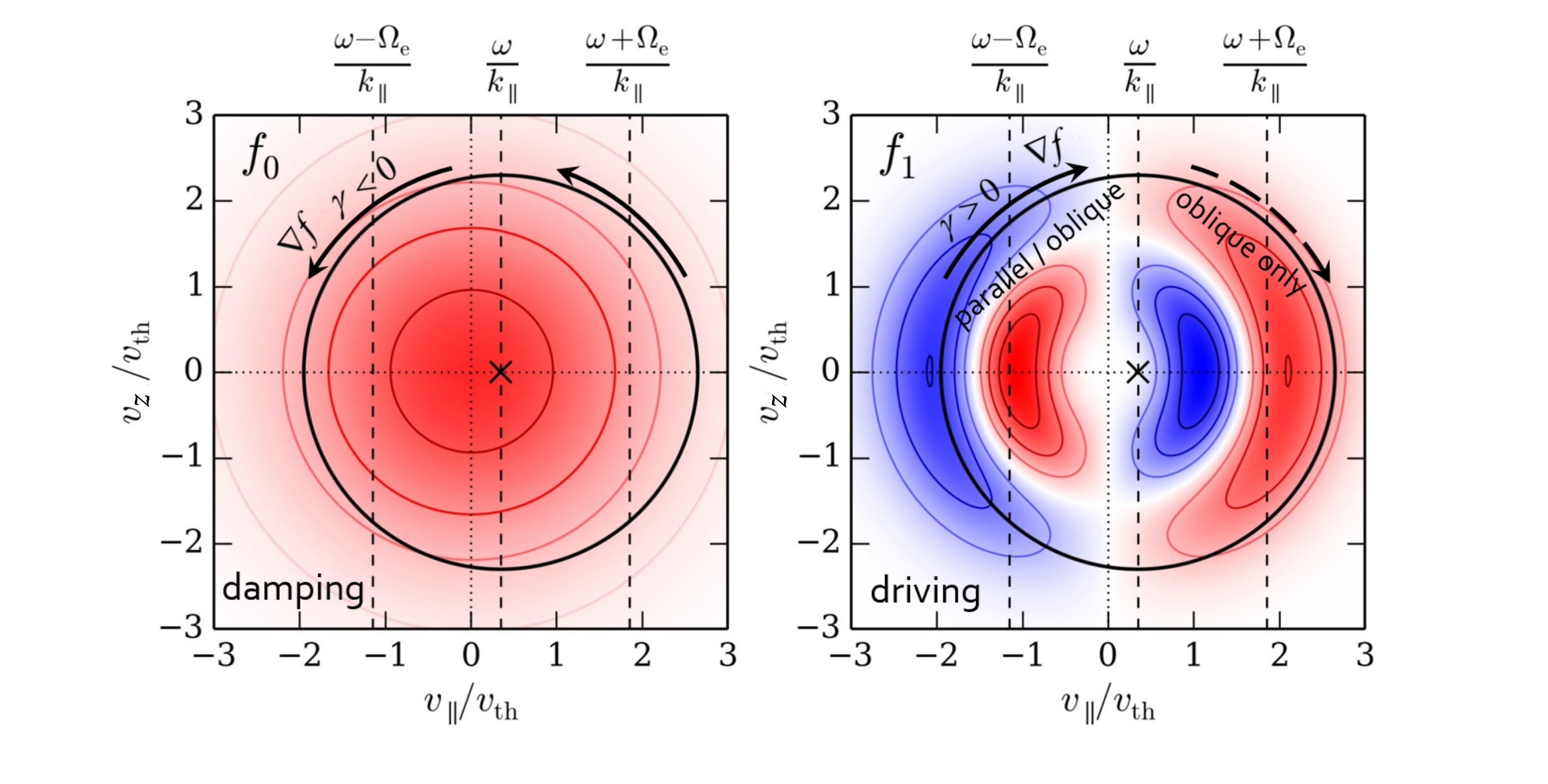}
\caption{An illustration of the mechanism of the whistler instability and marginality of the electron 
distribution function. The colored contours show Maxwell's distribution 
on the left and the anisotropic perturbation associated with a heat flux on the right 
(the heat flux is along the $v_{\parallel}$ axis). The left and right dashed vertical lines in 
both panels indicate the positions of the gyroresonances, while the central dashed lines 
correspond to the parallel phase velocity of a whistler. The solid circles 
demonstrate the contours of constant energy in the frame moving with the parallel phase 
velocity of the wave. We choose to use $v_z$ instead of $v_{\perp}=(v_y^2+v_z^2)^{1/2}$ for the vertical axis solely because it is allowed to be negative, which makes for a more natural visual representation of the distribution function. The instability grows when the distribution function near 
the resonances increases in the clockwise direction (for $v_z>0$) along the solid circles, as for 
the heat-flux perturbation on the right. Driving by the anisotropy is balanced 
by cyclotron damping on the bulk of isotropic particles (left panel). If the wave spectrum is broad enough ($\Delta k_{\parallel}/k_{\parallel}\sim 1$), 
electrons are scattered within a wide range of parallel velocities, and marginality is 
reached when electrons become isotropic in the wave frame. Both negative and positive 
resonances are only enabled for oblique whistlers, as described in the text.}
\label{fig:margin}
\end{figure*} 

A parallel heat flux introduces an asymmetry in the electron distribution function, and the
perturbed distribution can be expanded in small parameter $\epsilon = \mfp/L_T$, 
where $\mfp$ is the electron mean free path (either classical Spitzer or one associated with scattering off magnetic fluctuations) and $L_T$ is the scale length of the temperature 
gradient. The heat flux is 
\beq
q_{\parallel} \sim n \mfp \vth
\nabla T \sim \epsilon n \vth T \propto \epsilon,
\eeq
where $n$ and $T$ are the electron density 
and temperature. 

The electron distribution function is 
\beq
f(v,\xi) = f_0(v) + \epsilon f_1(v,\xi),
\eeq
where $f_0(v)$ is the unperturbed isotropic Maxwell distribution and $f_1(v,\xi)$ is 
an anisotropic perturbation that depends on $\xi$, the cosine of the electron pitch angle. 
Both of the components are shown in \figref{fig:margin} with equi-energy contours superimposed. 
For $f_1$ we use the shape of distortion that arises from the Knudsen expansion of the Boltzman equation 
with the simplest Krook operator describing isotropic collisions \citep[e.g.,][]{LE1992}:
\beq
\label{eq:Knuds}
f_1(v,\xi) = \frac{\xi v}{2\vth} \left ( \frac{v^2}{\vth^2} - 5 \right ) f_0(v).
\eeq
This is done only 
for illustrative purposes, and in fact any perturbation by a heat flux, with the proviso that no electron current is produced, can be used 
instead. All such perturbations will have similar features, namely, an excess of parallel momentum 
in the direction of the heat flux at $\sim 2 \vth$, its deficiency in the opposite direction, and a backflow 
of colder particles (the central dipole-shaped pattern in the right panel of \figref{fig:margin}) to 
cancel the electron current.     
\Figref{fig:margin} demonstrates that the Maxwellian part (left panel) absorbs energy from the wave [see 
\exref{eq:gammacircle}], while the anisotropy associated with the heat flux provides the 
free-energy source driving the instability.  To make it clearer, we can estimate the parallel phase speed 
of resonant whistlers in \exref{eq:gammapart}.  By taking $|v_{\parallel r}|\sim \vth$ (see the resonances 
in \figref{fig:margin}) and, therefore, $k_{\parallel}\sim \rho_e^{-1}$, we get
\beq 
\omega/k_{\parallel}= k_{\parallel} c^2 \Omega_e / \omega_p^2 \sim \vth/\beta_e.
\eeq
Let us also change variables in \exref{eq:gammapart} to $(v, \xi)$ and use the fact that 
$\omega \ll |k_{\parallel} v_{\parallel}|$. This leads to
\beq
\Gamma_s(\vc{v},\vc{k}) \propto \left ( \frac{\vth}{\beta_e}\frac{\dd}{\dd v} + \frac{\dd}{\dd \xi} \right )
f(v,\xi) \vert_{v_{\parallel}=v_{\parallel r}} 
\approx \left ( \frac{\vth}{\beta_e}\frac{\dd f_0}{\dd v} + 
\epsilon \frac{\dd f_1}{\dd \xi} \right )_{v_{\parallel}=v_{\parallel r}}. 
\label{eq:gammavmu}
\eeq
It is manifest now that the growth rate consists of two terms: the damping term that describes suppression of waves 
by the bulk of isotropic particles (electron cyclotron damping), and the driving term proportional 
to the anisotropic distortion of the distribution function, i.e., to the heat flux. Note that equation~(\ref{eq:gammavmu}) is written 
for $\kpl>0$; for $\kpl<0$ there is no instability because $\Gamma_s$ reverses its sign 
in \exref{eq:gammacircle}, and the driving term in \exref{eq:gammavmu} becomes negative, while the 
damping term proportional to the phase speed remains negative because the phase speed also changes its sign.

By estimating $\dd f_0 / \dd v \sim - f_0/\vth$, 
$\dd f_1/\dd \xi \sim f_1$, $f_1 \sim f_0$ and demanding marginal stability $\gamma=0$, we get
\beq
\label{eq:epsbeta}
\epsilon \beta_e \sim 1.
\eeq
Thus, the marginal heat flux is
\beq
\label{eq:margheatflux}
q_{\parallel}^{\rm m} \sim \beta_e^{-1} n \vth T.
\eeq
We have achieved the result anticipated in equation~(\ref{eq:genheatflux}): 
the marginal heat flux is limited by the value of the phase speed of resonant whistlers and, therefore, 
is fully controlled by the electron plasma $\beta_e$.

It is helpful for further discussion to go back and estimate the order of magnitude of the driving and damping terms, as we have so far been interested only in the sign of the growth rate.  Ignoring the angular dependence, $\mathcal{W}(\vc{p},\vc{k})$ in \exref{eq:gammak} can be estimated as $\mathcal{W}(\vc{v},\vc{k}) \sim m_e^2 v^3 / \hbar n$ [see \citealt{Melrose1980}, equation (7.46)]. Then, using equations~(\ref{eq:gammak}), (\ref{eq:Gamma}), (\ref{eq:gammavmu}), and the resonance condition $\kpl \sim \rho_e^{-1}$, we get
\bea
\label{eq:grow}
\gamma_{\rm grow}/\Omega_e &\sim& \epsilon,\\
\label{eq:damp}
\gamma_{\rm damp}/\Omega_e &\sim& 1/\beta_e. 
\eea

\subsection{Resonant wave-particle interaction: need for oblique whistlers} 

\label{sec:oblique}       
   
So far we have completely ignored the details of the resonant interaction 
between the heat-carrying electrons and whistler waves. Namely, we have simply considered that both gyroresonances 
are active, and particles with both negative and positive parallel velocities are 
scattered by the magnetic perturbations. However, because the whistler wave is an electromagnetic wave modified by 
gyrating electrons, it is right-hand polarized. Whistlers 
that propagate along the mean magnetic field have a right-hand circular polarization. This means 
that they strongly interact only with electrons moving opposite to the wave (and the field) because 
those are the ones that co-rotate with the electric-field vector of the wave in the frame moving 
with the parallel electron velocity $v_{\parallel}$. The corresponding resonance condition 
for parallel propagation is $\omega-k_{\parallel} v_{\parallel}\approx -k_{\parallel} v_{\parallel} = \Omega_e$, where 
the positive sign before $\Omega_e$ is fixed by the right-hand polarization of the wave. This 
is the left gyroresonance in \figref{fig:margin}. Analysis of the linear whistler growth 
rate done by substitution of the whistler dispersion properties into the function $\mathcal{W}
(\vc{p},\vc{k})$ in \exref{eq:gammak} and using the Knudsen expansion of the electron distribution 
function in the presence of a small collisional heat flux (set by isotropic Coulomb collisions) is 
a difficult task for whistlers with arbitrary propagation angles. It is drastically simplified 
for the case of near-parallel propagation, and predicts that the maximum growth rate is reached 
for strictly parallel whistlers that resonate, as we have noted, with electrons moving 
opposite to the heat flux (see PE98). Such whistlers are not expected to scatter the 
heat-carrying electrons. 

From the right panel of \figref{fig:margin}, it can be seen that 
it is regions of high $v_{\perp}$ and negative $v_{\parallel}$ that drive the parallel whistlers. Scattering off the parallel 
modes modifies the electron distribution, and it evolves to a new current-free distribution, different from the initial state obtained from the Knudsen expansion.   
Because such scattering is not isotropic, the 
dependence of the new marginally stable distribution function 
on the cosine of the electron pitch angle $\xi$ no longer has to be in the simple dipole form $f_1=\xi \phi_1(v)$ as in the Knudsen expansion [see equation~(\ref{eq:Knuds}) and \figref{fig:margin}].  
The new state may be characterized by more depletion of 
the anisotropy at negative $v_{\parallel}\sim -2\vth$ 
compared with the one at $v_{\parallel}\sim 2\vth$ associated 
with the heat flux. We will further show in \secref{sec:distrsim} that such state is indeed seen in our numerical simulations. In this state, the maximum growth rate can be achieved for oblique whistler propagation instead. Oblique modes, in contrast 
with parallel, are right-hand {\it elliptically} polarized, 
which can be represented as a combination of right- and left-hand circularly polarized waves. Thus, both 
positive and negative resonances, $k_{\parallel} v_{\parallel}=\pm \Omega_e$, become 
active (see \figref{fig:margin}), and efficient scattering of the heat-carrying particles is possible.  
PE98 used quasilinear equations to predict that the final marginal state is indeed 
characterized by whistlers propagating at a large angle to the mean magnetic field. 
However, their result is not fully self-consistent because they still relied on the 
approximation of the whistler dispersion relation for near-parallel propagation, while 
the resulting angle of propagation was found to be large.
In the face of significant analytical complexity of even the quasilinear treatment 
of the instability, numerical simulations are vital in order to test the expectation that oblique whistlers will dominate 
the marginal state.

\subsection{Saturated magnetic field}

\label{sec:satmf}

Let us assume that electron orbits are only weakly perturbed by the unstable 
whistlers, nonlinear effects can be neglected due to the smallness of the 
saturated magnetic fluctuations, and saturation is quasilinear (an assumption that will be confirmed in \secref{sec:scattsat}). 
The scattering rate $\nu_{\rm scatt}$ of resonant electrons
can be expressed via $\epsilon$ (see the beginning of \secref{sec:margheatflux})
as 
\beq
\label{eq:nuscatteps}
\nu_{\rm scatt} = \frac{\vth}{\mfp} = \epsilon^{-1}\Omega_e \frac{\rho_e}{L_T} ,
\eeq
or, using the marginality condition \exref{eq:epsbeta},
\beq
\label{eq:nuscatt}
\frac{\nu_{\rm scatt}}{\Omega_e}  \sim \beta_e \frac{\rho_e}{L_T}.
\eeq  
Given that the resonant magnetic perturbations arise at the electron Larmor 
scale, both gyroresonances are available (assuming oblique propagation), and 
the whistler spectrum is sufficiently broad to scatter particles isotropically, 
we can estimate the effective pitch-angle scattering rate from Bohm diffusion:
\beq
\label{eq:bohm}
\frac{\nu_{\rm scatt}}{\Omega_e}  \sim \frac{\delta B^2}{B_0^2} ,
\eeq
where $\delta B$ is the saturated magnitude of the magnetic perturbations 
at the resonant parallel wavelength $k_{\parallel {\rm r}} = \Omega_e/\xi v$ and $B_0$ the mean magnetic field. 
This allows one to obtain the saturated magnetic field:
\beq
\label{eq:satfield}
\frac{\delta B^2}{B_0^2} \sim \beta_e \frac{\rho_e}{L_T}. 
\eeq
Note that this quasilinear saturation level is extremely low for astrophysical plasmas. For galaxy clusters, $\rho_e/L_T \lesssim 10^{-13}$ at temperature $T=10$ KeV, magnetic field $B_0=1~{\rm \mu G}$, and $L_T\gtrsim 10$ kpc, while $\beta_e\sim100$. The resulting saturation level then is $\delta B^2/B_0^2 \sim 10^{-11}$. 

\section{Numerical simulations}
\label{sec:num}
By performing numerical simulations with different $\beta_e$ and $\rho_e/L_T$, we will now check the validity of our assumptions and qualitative results: 
the oblique propagation of whistlers, the expressions for 
the marginal heat flux \exref{eq:margheatflux}, effective pitch-angle scattering 
rate \exref{eq:nuscatt}, and the saturated level of magnetic fluctuations \exref{eq:satfield}, all as functions of $\beta_e$ and $\rho_e/L_T$. 

 
\subsection{Numerical setup}
We use the relativistic electromagnetic particle-in-cell code {\it TRISTAN} (\citealt{Buneman1993}, \citealt{Spitk2005}). 
Our simulation domain is an elongated 2D grid of size $N_x \times N_y = 2560 \times 512$ 
with the number of particles per cell varying from 200 to 500 depending on the run to obtain convergence and an acceptable signal-to-noise ratio. 
The simulation is 2.5D, meaning that particle velocities are 3D. The ions 
are motionless and form a charge-neutralizing background (see, however, \secref{sec:ions} 
on the role of ions). The mean magnetic field 
$\vc{B_0}$ points in the positive $x$ direction, while the initial temperature gradient 
is set in the negative $x$ direction. 
The grid size measured in the electron Larmor radii  $\rho_{e1}$ at the left (hot) end of the box is fixed at 
$L_x \times L_y \approx 125 \times 25 \rho_{e1}$, and $\rho_{e1}\approx 20$ cells. 
The electron temperature at the hot wall is such that 
the ratio of the thermal electron speed to the speed of light is $v_{{\rm th}1} / c \approx 0.33$. We vary the initial electron plasma beta 
$\beta_e =8 \pi n T / B_0^2=$(10,15,25,40). For the run with $\beta_e=40$, we increase the 
grid resolution by a factor of 1.5 to better resolve the electron skin depth $d_e$. At the hot wall, where the particle density is the smallest (see below), $d_{e1}\approx$(6.5,5.4,4.2,4.9) cells for the corresponding range of $\beta_e$, which gives $\rho_{e1}/d_{e1}\approx$(3.1,3.7,4.8,6.1). 

For the electrons' initial conditions in velocity space, we employ the isotropic 
Maxwell distribution with temperature $T_0(x)$ that decreases linearly from $T_1$ to $T_2$ 
along $x$. $T_1$ is kept constant in all the runs, while $T_2$ is varied. We vary 
$T_1/T_2=(1.5,2,3)$ in runs with the same $\beta_e=15$, while 
the scan over $\beta_e$ is performed at the same $T_1/T_2=2$. 
The mean electron density $n_0$ is distributed so as to keep the electron pressure $p$ uniform across 
the box, i.e., $n_0(x)\propto1/T_0(x)$, and does not evolve in time (the electrons are 
bound to the ions by quasineutrality). This way, the initial distribution is close
to what we expect to observe when the instability saturates by scattering  
and isotropizing the electrons, making the plasma effectively collisional. While the temperature gradient evolves in the simulation and is not strictly linear at saturation, setting the particle density  $n_0(x)\propto1/T_0(x)$ largely reduces spatial variations of pressure and, consequently, $\beta_e$. Alternatively, 
we could have used a collisionless initial condition with uniform density and counterstreaming 
electrons at temperatures $T_1$ and $T_2$, represented in velocity space by two ('hotward' and 'coldward') 
Maxwellian hemispheres with densities chosen so that the net electron current is zero. 
In this case, electron scattering would have led to formation of an additional mean 
electric field to compensate for the temperature (and, due to uniform density, pressure) 
gradient. We have tried both types of initial conditions and observed no significant 
differences in the properties of the saturated state, and thus use the former way 
in what follows.

\begin{figure*}
\centering
\begin{minipage}[b][0.5\textheight][s]{0.99\textwidth}
  \centering
  \includegraphics[width=\textwidth]{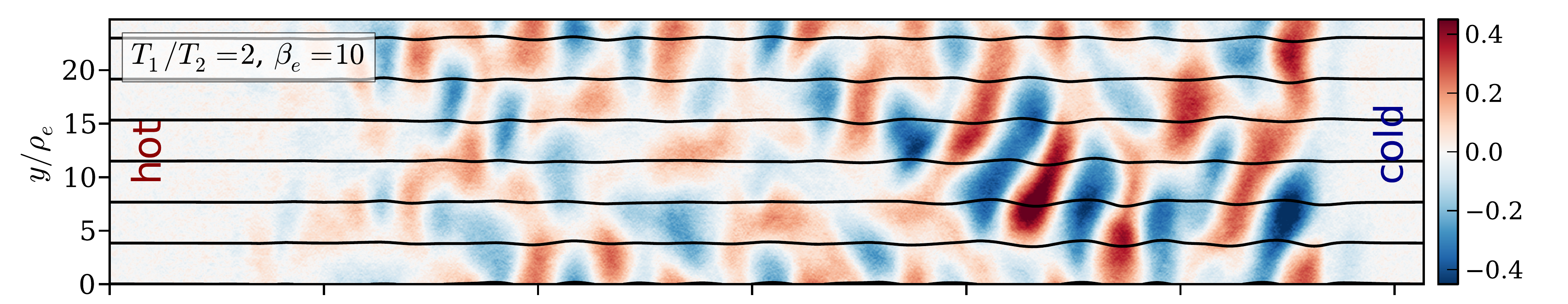}
  \vfill
  \includegraphics[width=\textwidth]{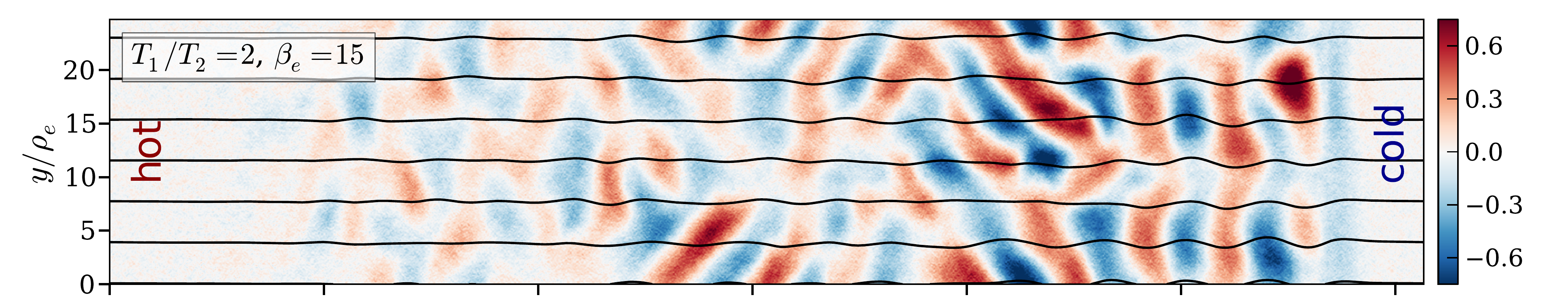}
  \vfill
  \includegraphics[width=\textwidth]{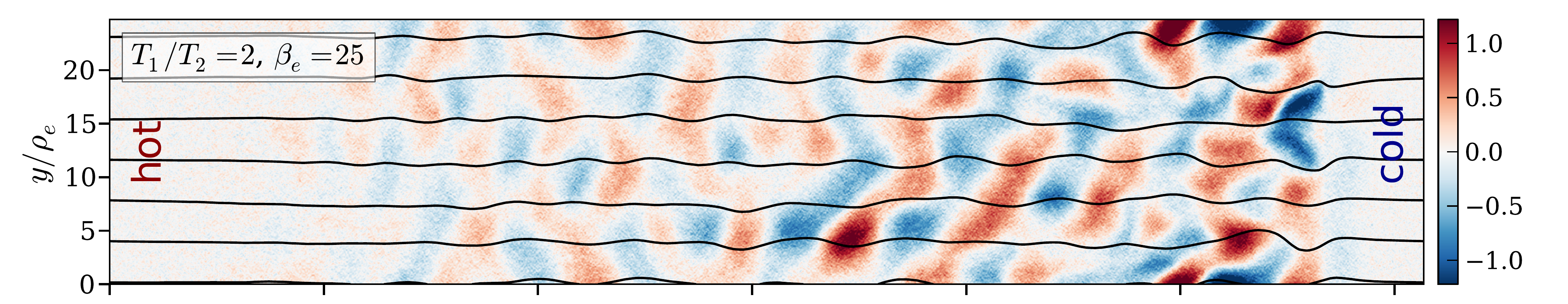}
  \vfill
  \includegraphics[width=\textwidth]{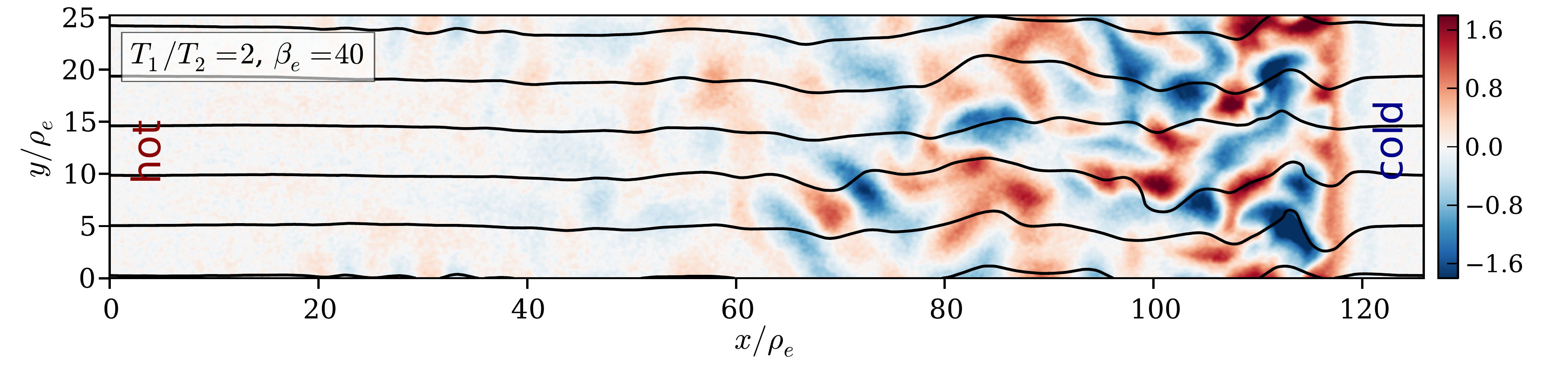}
\end{minipage}
\caption{The spatial structure of the $z$-component of the magnetic field generated by the heat-flux-induced 
  whistler instability. The magnetic-field lines are shown as the black contours. The temperature gradient points opposite to the $x$ axis, i.e., the 
  heat flux is in the positive direction. The whistlers propagate in the direction of the heat flux 
  from the hot (left) to the cold (right) wall, which are located $\approx 10 \rho_{e1}$ away from the edges 
  of the box. The regions behind the walls are used to dissipate the energy of the incident 
  waves. The saturated state of the instability is characterized by oblique whistler 
  propagation in a wide range of angles. }
\label{fig:simB}
\end{figure*} 


Proper boundary conditions are essential for simulations of an instability constantly 
driven by a sustained temperature gradient. For the particles, we use periodic 
boundary conditions along $y$ and reflective boundary conditions along $x$. A 
particle reflected from a wall acquires a random velocity drawn from the velocity distribution function $f_0(\vc{v},T_{1,2})v_x$, where $f_0$ is Maxwell's distribution at temperature $T_{1,2}$. This ensures that the incoming flux of particles colliding with a 
wall is equal to the flux of the reflected particles with new velocities. The reflected flux corresponds 
to a flow of Maxwellian particles at temperature $T_{1,2}$ from an infinite half-space through the plane of a wall.

For the electric and magnetic fields, we introduced absorbing boundary conditions to reduce 
reflection of the wave modes generated by the instability from the walls. We put thermal 
reservoirs of particles behind the walls. The reservoirs have width $L_D=192$ cells 
($\sim$ the typical wavelength of whistlers, i.e., several electron Larmor radii $\rho_{e1}$). 
In the reservoirs, which serve as absorbers, the fields $\vc{B}$ and $\vc{E}$ are evolved to decay as \citep{Umeda2001}
\bea
\vc{B}^{n+1/2}(x) &=& {\alpha}_M(x) \{ \vc{B}^{n-1/2}(x) - c \Delta t \nabla \times \vc{E}^n (x)\},\\
\vc{E}^{n+1/2}(x) &=& {\alpha}_M(x) \{\vc{E}^{n-1/2}(x) - \Delta t [4 \pi \vc{j}^n (x) - c \nabla \times \vc{B}^n (x)]\},
\eea
where $\vc{j}$ is the current density, $c$ the speed of light, and $\Delta t$ the 
timestep. 
The masking parameter  ${\alpha}_M(x)<1$ gradually decreases into the reservoirs in 
order to avoid numerical reflections at the absorber boundary:
\beq
{\alpha}_M(x) = 1 - \left ( r \frac{|x-x_{1,2}|}{L_D} \right )^2,
\eeq
where $x_{1,2}$ are the walls' positions. The parameter $r\approx0.02$ regulates 
the gradient of the masking function and is adjusted to utilize the width of the 
absorbing regions most effectively for a given group velocity of the waves 
\citep{Umeda2001}. One must put particles in the absorbers to reduce wave reflection by matching the impedance of the absorbers with the one of the plasma in the main domain near the walls. This provides good absorption 
for perpendicular wave incidence. For off-axis waves, however, some reflection 
is still present as can be seen at higher wave amplitudes in \figref{fig:simB}.

\subsection{Results}

\subsubsection{Field structure}

\label{sec:fieldstruct}

In the absence of scattering, the initial isotropic electron distribution with a 
temperature gradient in the negative $x$ direction is unstable to generation of whistlers propagating with a group velocity $v_{\rm g} \sim \vth / \beta_e$ opposite to the 
temperature gradient, i.e., in the direction of the heat flux. The unstable right-hand polarized modes 
grow at the scale of $\sim 10 \rho_e$, practically independent of $\beta_e$ and $T_1/T_2$
\footnote{Linear theory predicts a very weak, $k^{\rm max}_{\parallel} \rho_e \sim (\epsilon \beta_e)^{1/6}$, dependence of the 
wave number of maximum growth on $\beta_e$ and $\epsilon=\mfp/L_T$, where $\mfp$ here is set by isotropic Coulomb collisions (see \citealt{LE1992}, 
PE98)}. 
The instability saturates in the state shown in \figref{fig:simB}. The 
presence of oblique modes in the final state is manifest. 

\begin{figure}
\centering
\minipage[t][][t]{0.46\textwidth}
\centering
  \includegraphics[scale=0.33]{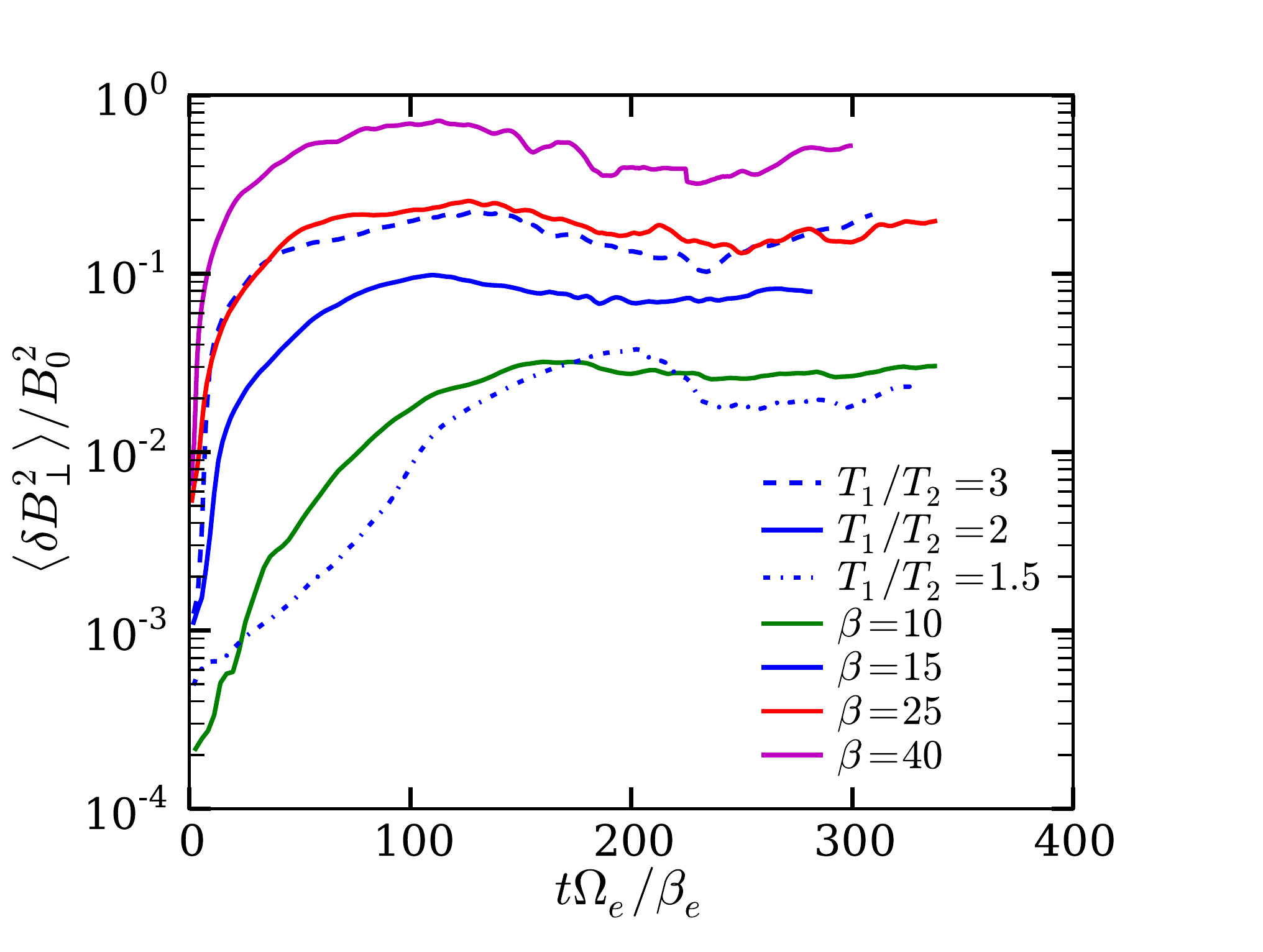}
  \caption{Evolution of the mean perpendicular magnetic-energy density in 
						 whistler modes for runs with different $\beta_e$ and $T_1/T_2$.}
  \label{fig:simBev}
\endminipage\hfill
\minipage[t][][t]{0.46\textwidth}
\centering  
\includegraphics[scale=0.33]{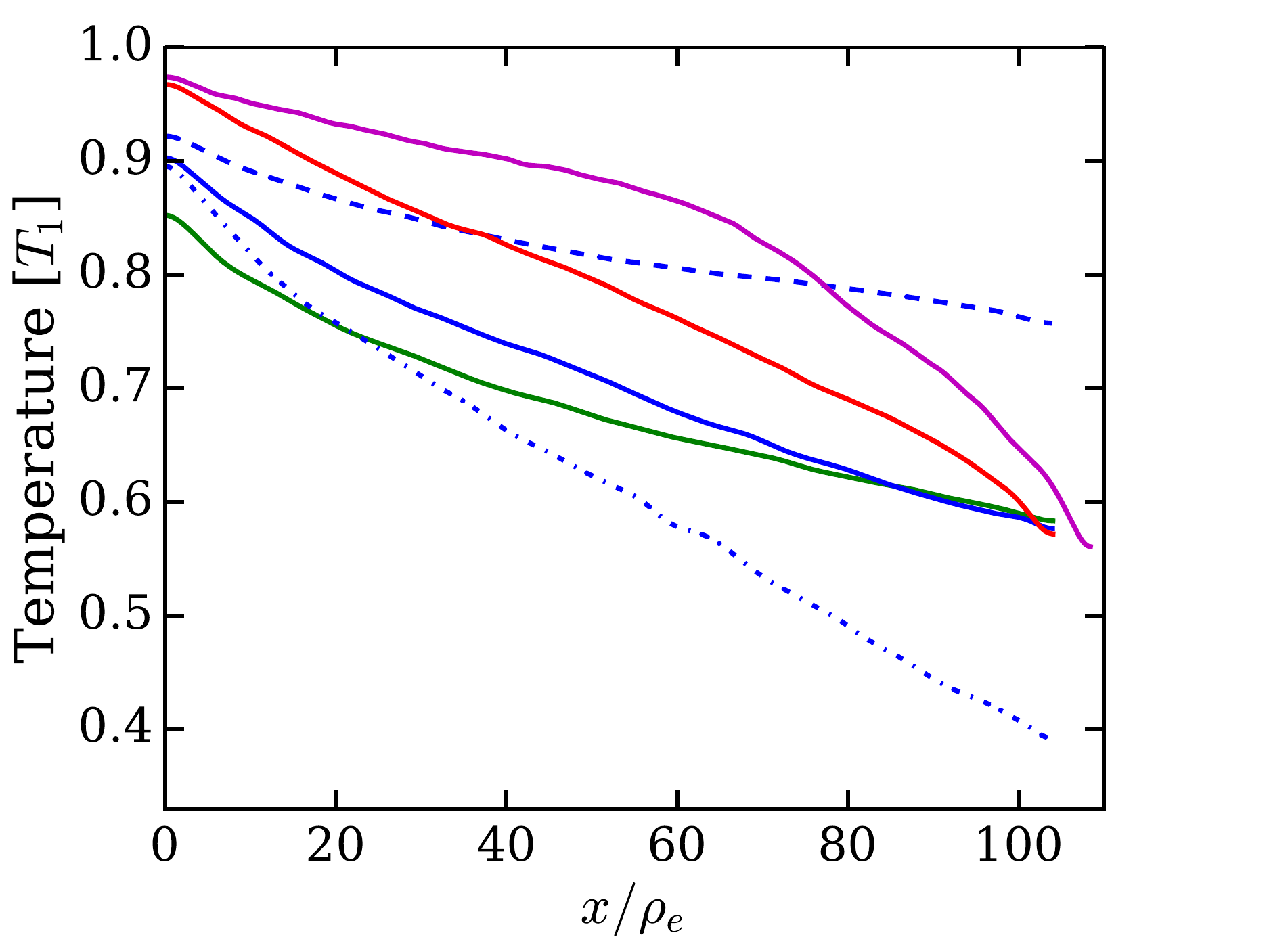}    
\caption{Temperature profiles normalized by the temperature $T_1$ of the hot wall at saturation.}
\label{fig:T}
\endminipage
\end{figure}

The waves grow as they travel away from the 
hot wall and reach full saturation in the right half of the box.  We plot the evolution 
of the mean magnetic-energy density in the perpendicular component of the magnetic field in 
\figref{fig:simBev}. The time required by the instability to saturate completely is rather long, 
only a few times smaller than 
the time it takes 
the waves to cross the box, $t_{\rm cross}  \Omega_e  \sim  L_x \beta_e / \rho_e \sim$ 1000-4000. 
This leads to the magnetic perturbations at saturation noticeably growing along $x$, reaching a 
plateau past the center of the box. The final amplitude of these perturbations can be determined purely by 
quasilinear saturation only where the field becomes spatially homogeneous, whereas before that,
wave advection likely plays a significant role, removing energy before quasilinear saturation fully comes into play. 
This is a consequence of non-periodic boundary conditions, and to alleviate the problem, 
one can either increase the box size, or simply calculate all the important quantities required 
to check theoretical assumptions in 
regions where the magnetic field is spatially homogeneous. Because the former route is quite 
computationally expensive, we have chosen the latter. We shall see that this does allow us to 
confirm our predictions. 

\begin{figure}
\centering
\begin{tabular}{c c c c}    
     \includegraphics[scale=0.34]{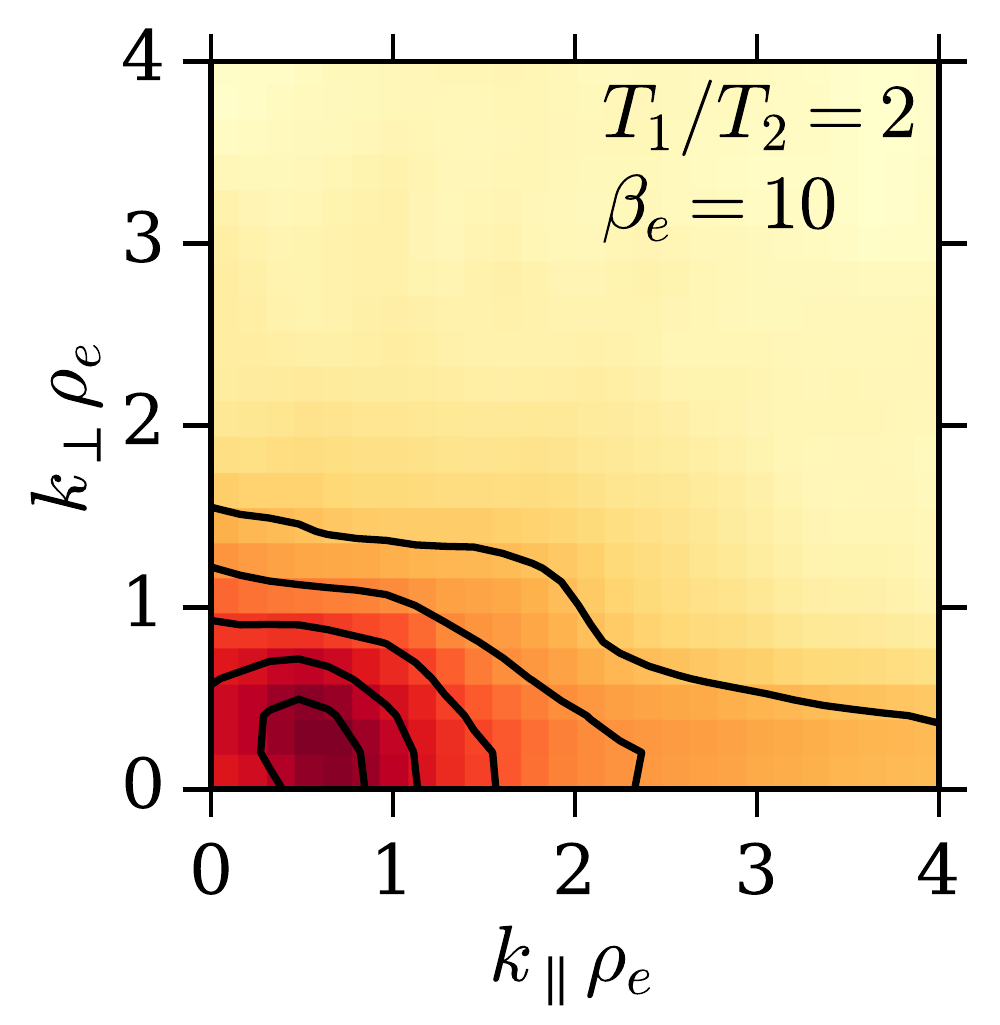} &
     \includegraphics[scale=0.34]{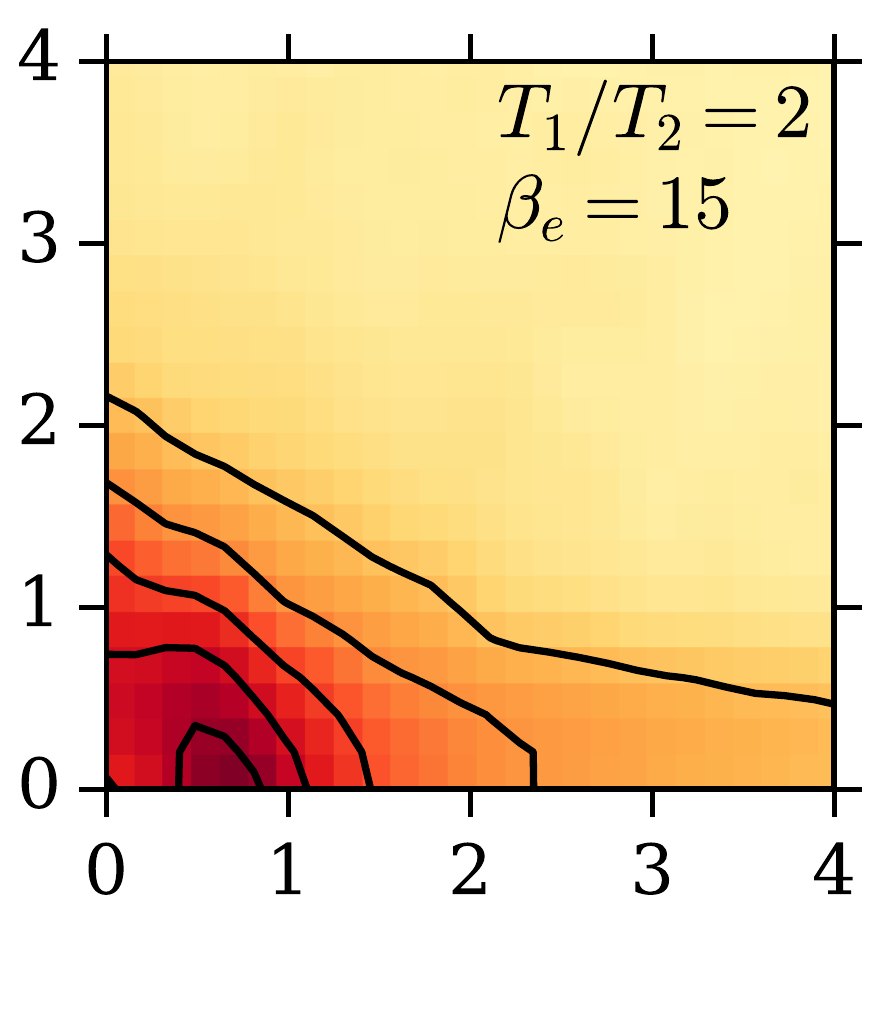} &
     \includegraphics[scale=0.34]{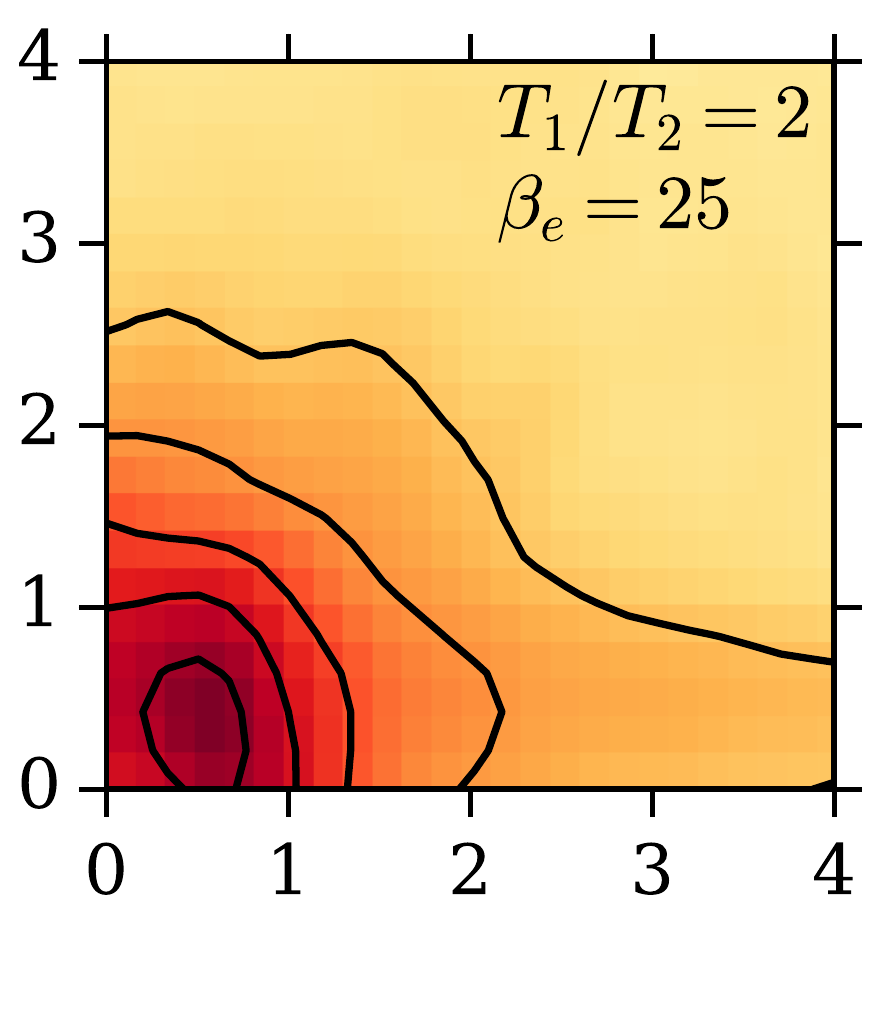} &
     \includegraphics[scale=0.34]{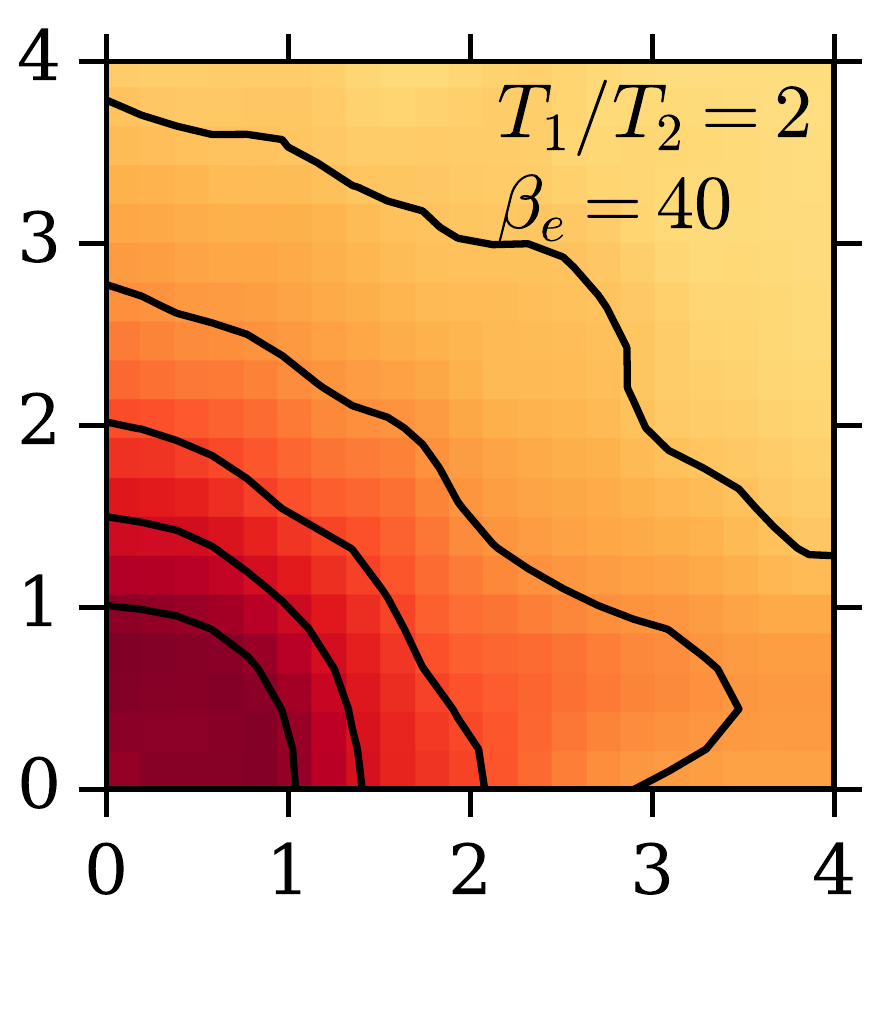}        
\end{tabular}
\caption{The logarithm of the 2D power spectrum of the $z$ component of the magnetic field 
  produced by the whistler instability for different values of $\beta_e$ and fixed $T_1/T_2=2$. 
  The contours correspond to logarithmic increments $\approx2.4$. The spectrum has been 
  calculated for the right third of the box, where the field power is even along $x$ in all the runs, and averaged over several time snapshots.
  The spectrum peaks at $\kpl \rho_e\sim 0.7$, largely independently of $\beta_e$, 
  as predicted by the linear theory.  The magnetic energy is distributed over a broad range of angles 
  rather than being concentrated along the parallel direction, 
  thus potentially allowing effective scattering of particles propagating at any 
  angles.}
  \label{fig:simPS}  
\end{figure}

The 2D power spectrum of the $z$ component (perpendicular to the picture plane in 
\figref{fig:simB}) of the magnetic field on scales larger than the electron Larmor radius 
is shown in \figref{fig:simPS} for runs with different $\beta_e$. 
The power spectrum peaks at ${\kpl}^{\rm max} \rho_e 
\approx 0.7$, corresponding to the scale of $\approx 10 \rho_e$, mostly independently of 
$\beta_e$. There is substantial power at all propagation angles, which confirms the hypothesis that the 
saturated marginal state is not restricted to parallel whistlers (as  
conjectured by \citealt{LE1992} and shown using quasilinear theory by PE98). The width of the excited spectrum 
of whistlers can be seen to be $\Delta \kpl/{\kpl}^{\rm max} \sim 1$.

The profiles of electron temperature in the saturated state are shown in  \figref{fig:T}. They are seen to 
be close to linear, with the exception of the cases with the lowest and the highest beta, thus producing only minor 
variations of electron pressure over the box. The difference between the calculated  temperature near 
the walls and the temperatures of the walls (and of the absorbing regions) is caused by the anisotropy of the electron 
distribution function: a larger anisotropy produces a bigger difference between the temperature of 
particles moving away from the wall with new thermal velocities and the temperature calculated by averaging 
over all directions of particle motion (as in the profiles).  Larger $\beta_e$ lead to 
stronger magnetic perturbations, therefore more isotropic electron distribution and smaller temperature 
difference between the plasma and the walls.   

\subsubsection{Marginal electron distribution}
\label{sec:distrsim}

\begin{figure}
\centering
  \begin{tabular}{@{}l@{}@{}l@{}}
    \includegraphics[width=67mm]{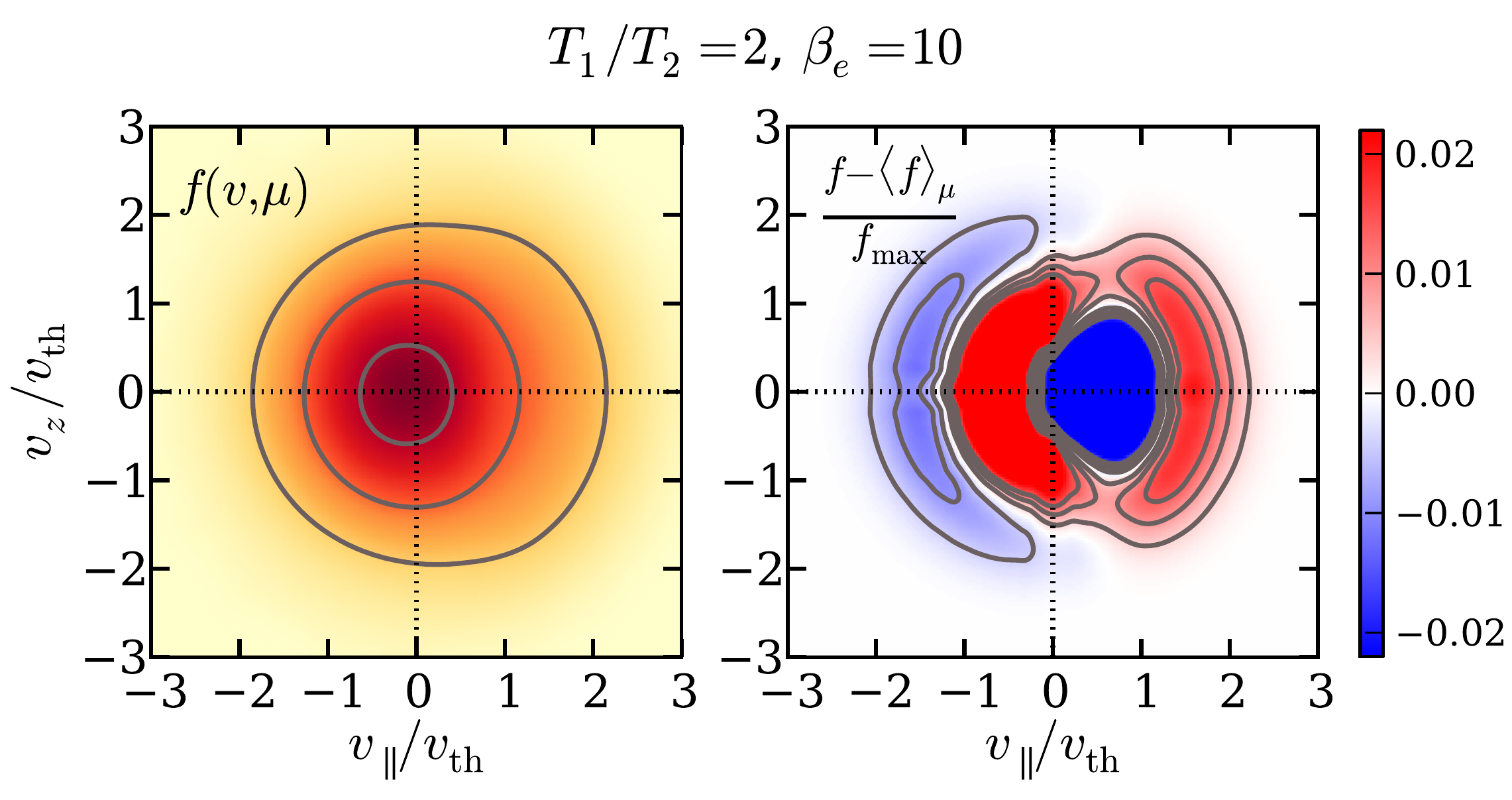} &
    \includegraphics[width=67mm]{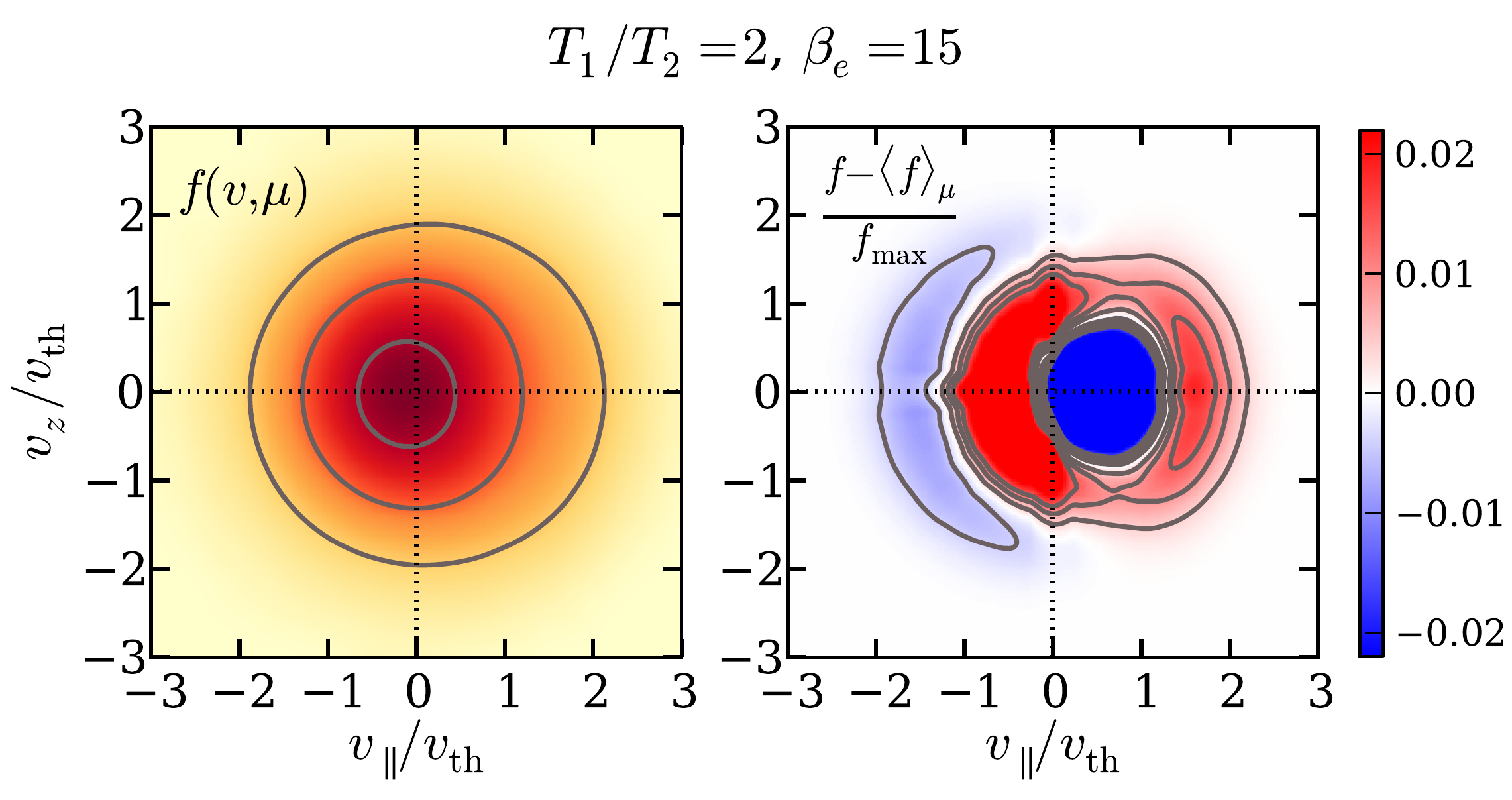}\\
    \includegraphics[width=67mm]{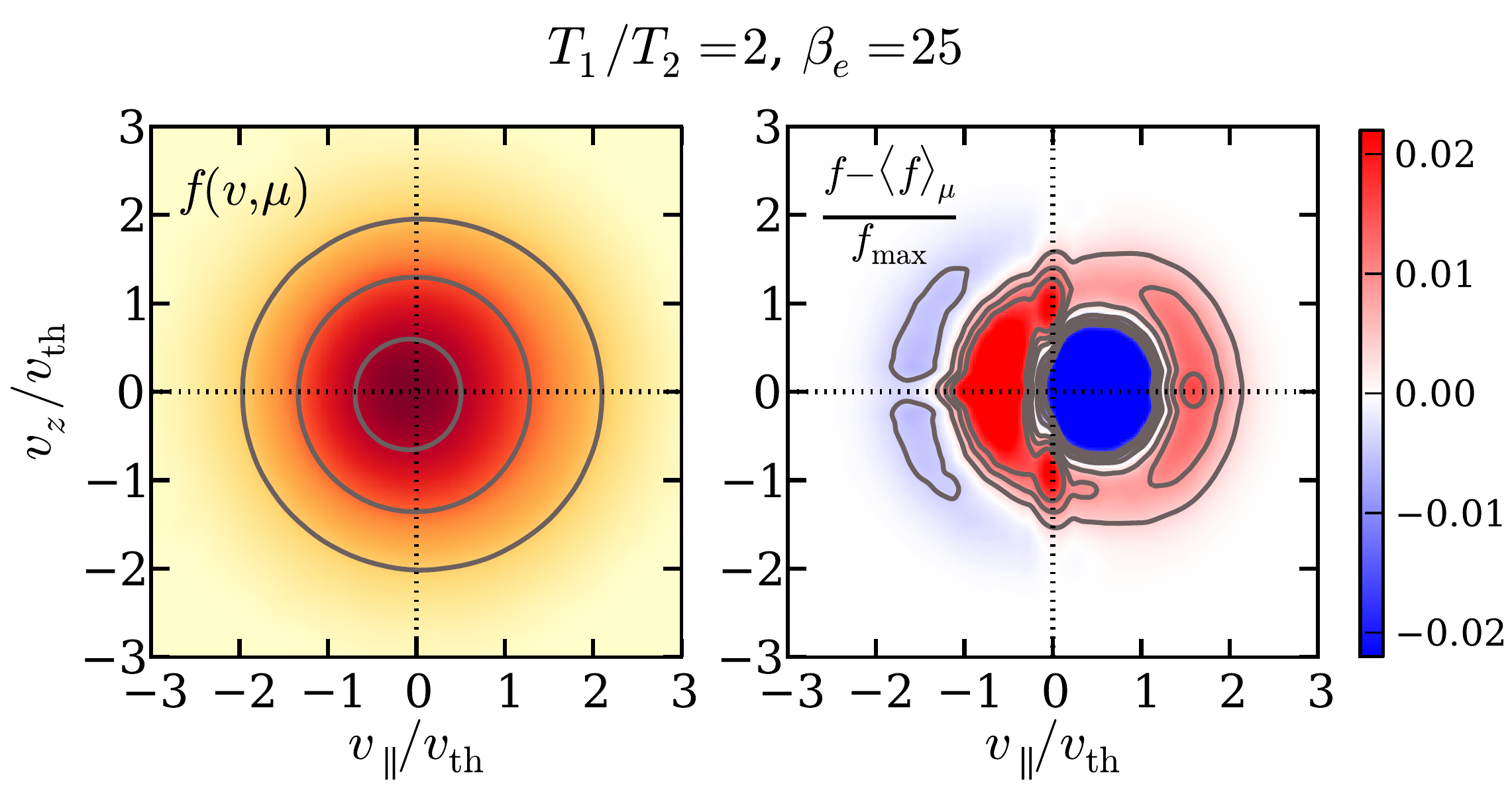} &   
    \includegraphics[width=67mm]{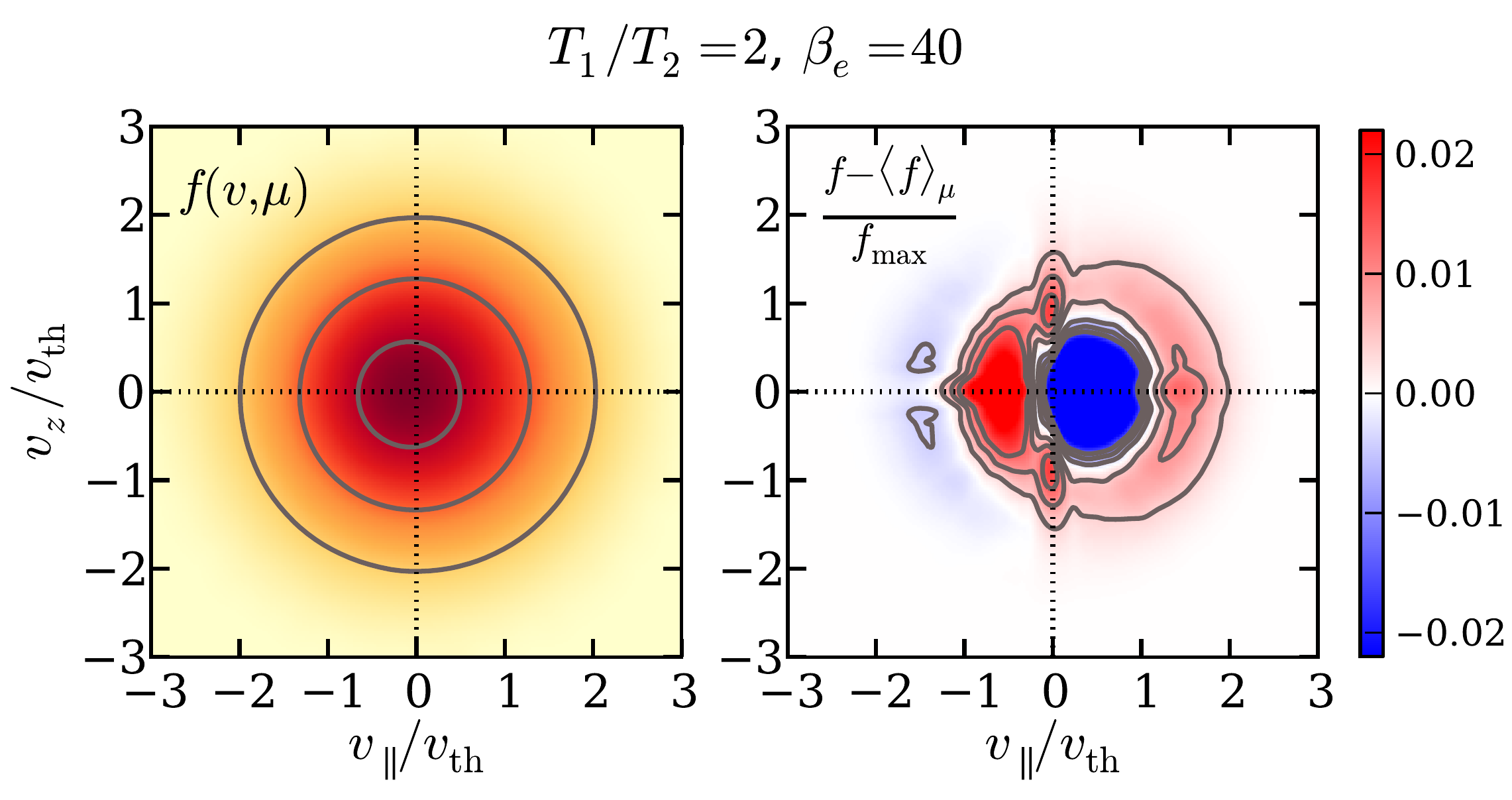}

  \end{tabular}
  \caption{The marginal electron distribution functions obtained from the
  simulations. The left column shows the total distribution 
  function and demonstrates the isotropization of the distribution at higher $\beta_e$. 
  The right column shows the anisotropic part of the distribution function 
  driven by the heat flux. Depletion of particles with negative (hotward) parallel velocities 
  is clearly seen. }
\label{fig:simdf}
\end{figure}

Let us analyze the perturbed part of the electron distribution function 
in our simulations. To do 
so, we calculate the distribution function in the right third of the box, 
average it over electron pitch angles, and subtract the averaged 
part from the calculated distribution. We do so for several time snapshots and average 
the distribution functions over them.  In this manner, we obtain the perturbation 
associated with the heat flux and the instability. The shape of the perturbation 
is similar to the one we used as an illustrative example of the anisotropy 
produced by a heat flux in \secref{sec:margheatflux}, and is presented 
in \figref{fig:simdf} on the right, along with the total distribution function 
on the left. The distribution that we show here is a function of $v_x$ and $v_z$, i.e., 
it has been integrated over $v_y$. 
The central region of the perturbation ensures the current-free condition\footnote{It also 
provides additional damping at low resonant parallel velocities $\vpl\lesssim\vth$: the term 
associated with the heat-flux perturbation in \exref{eq:gammavmu} becomes negative and adds to 
the negative term associated with the cyclotron damping by the isotropic part of the distribution 
function. Such damping is balanced by driving at the same $\vpl$ but higher $\vpd$, where 
the $\xi$-derivative of the perturbation is positive.}. The outer 
parts, on the other hand, are those who drive the instability (recall 
\figref{fig:margin}). It is clear that larger $\beta_e$ reduce the overall 
anisotropy.

As we mentioned in \secref{sec:oblique}, for the dipole-shaped distribution 
resulting from the Knudsen expansion of the Boltzman equation with an isotropic 
collision operator, parallel whistlers grow the fastest, while they do not interact 
with the coldward (right) heat-flux-carrying part of the distribution. 
A modification of the perturbation caused by scattering off the unstable modes, 
however, can result in a state favorable for generation of oblique modes, which 
brings both positive and negative gyroresonances into action. 
This may happen if initially the dipole perturbation generates parallel 
whistlers that resonate with the hotward-streaming electrons, which depletes 
the hotward part of the distribution. Then the final state becomes 
germane to oblique whistlers that grow and resonate with the coldward 
electrons.
The asymmetry in the morphology of the perturbed distribution seen in \figref{fig:simdf} 
can be an indication of such a shift of marginal stability to oblique propagation\footnote{It can be shown using QLT under the approximation of small 
propagation angles that an asymmetry of the scattering operator leads to a 
non-zero propagation angle of the fastest growing whistlers (see PE1998).}. It is 
especially vivid at larger $\beta_e$: the hotward part of the distribution is 
significantly more depleted than the coldward part associated with the heat flux. 
This final state of the electron distribution function obtained in our simulations 
therefore can support the above picture.

\def\arraystretch{0}

\subsubsection{Heat flux}
\label{sec:heatflux}

Now we are in a position to examine whether the heat flux is indeed governed by 
the instability and limited by the marginal anisotropy. 
The fluxes are averaged over the computational domain (there is no systematic 
spatial variation of the heat flux at saturation, otherwise a build-up of energy would occur) 
and several time snapshots.
As we anticipate 
\begin{figure}
\centering  
    \includegraphics[width=74mm]{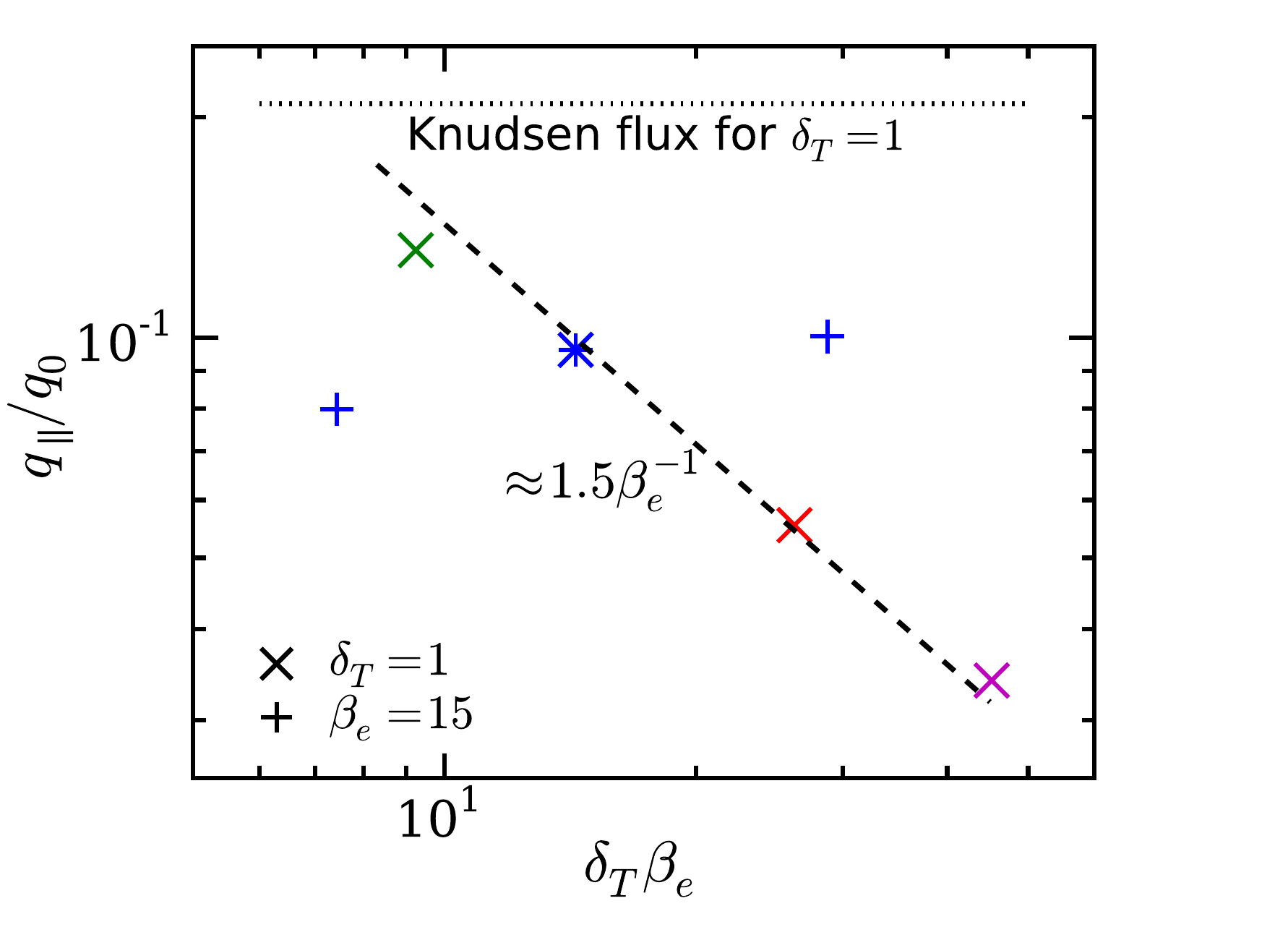}   
  \caption{The heat fluxes measured in the numerical simulations as functions 
  of $\delta_T \beta_e$, where $\delta_T = T_1/T_2-1$. The pluses 
  represent the runs with the same plasma beta $\beta_e=15$, while 
  the crosses of the same colors as in \figref{fig:simBev} show 
  the runs with the same temperature gradient $T_1/T_2=2$. We have made corrections for the small variation of the mean thermal pressure (and, therefore, the effective $\beta_e$) in different runs.}
\label{fig:simQ}
\end{figure}
suppression of the heat flux given by \exref{eq:margheatflux}, 
it is convenient to normalize the flux by $q_0=\langle m_e n \vth^3  \rangle \approx 
2 n_1 T_1 \langle v_{\rm th} \rangle$, where the angle brackets indicate averaging over the box, 
and we have taken the pressure $n T$ to be close to uniform across the
box. 

The heat flux as a function 
of $\delta_T \beta_e$, $\delta_T=T_1/T_2-1$, is shown in \figref{fig:simQ}. 
At constant $\beta_e=15$ but different 
$T_1/T_2$, there is only a small scatter in the heat flux as $\delta_T$ is varied by a factor of 4. This agrees with the qualitative 
expectation that the flux has no dependence on the imposed temperature gradient when 
the marginal state is reached (\secref{sec:general}).

The heat fluxes taken at 
constant $\delta_T$ and different $\beta_e$ are well fitted by $q_{\parallel}/q_0 \approx 1.5 \beta_e^{-1}$. 
We also show the collisionless Knudsen heat flux (also normalized by $q_0$) at the fixed $T_1/T_2=2$ 
to demonstrate the relative amount of suppression in the simulation. This heat flux corresponds to a velocity distribution composed of 
two Maxwellian hemispheres associated with two opposite electron fluxes from 
the respective walls with even electron density and no electric field. The Knudsen heat flux is the maximum flux attainable in any configuration with fixed $T_1$ and $T_2$, and is equal to
\beq
q_K = \frac{1}{\sqrt{\pi}} m_e n_0 \left [ \alpha v_{\rm th 1}^3 - (1-\alpha) v_{\rm th 2}^3 \right ],
\eeq 
where $n_0$ is the mean electron density (density is even in a completely collisionless plasma), 
$\alpha$ and $1-\alpha$ represent the fractions of particles moving in the coldward and hotward 
directions respectively in order to make for zero net electron current, viz.,
\beq
\alpha(T_1,T_2) = \left [  1 + \left ( \frac{T_1}{T_2} \right )^{1/2} \exp{\frac{T_1-T_2}{T_1 T_2} }  \right ]^{-1}.
\eeq  
We see that if our fit of the $\beta$ dependence is correct, no suppression 
below $\beta_e \sim 10$ should be present. This could be the reason why 
the data point at $\beta_e\approx10$ is a slight outlier.

Clearly, the simple argument in \secref{sec:phys} based on saturation in the marginal state of the whistler instability indeed leads to conclusions 
supported by numerical simulations. Thus, we expect the heat flux to be fully controlled 
by the instability and not being able to exceed its marginal value \exref{eq:margheatflux}: for the general case of a plasma 
with a small temperature gradient,
\beq
\label{eq:finalflux}
q_{\parallel}^{\rm max} \approx 1.5 \beta_e^{-1} n\vth^3.
\eeq

\subsubsection{Scattering rate and saturated magnetic field}
\label{sec:scattsat}

\begin{figure}
\centering 
\minipage[t][][t]{0.45\textwidth}
\centering
    \includegraphics[width=68mm]{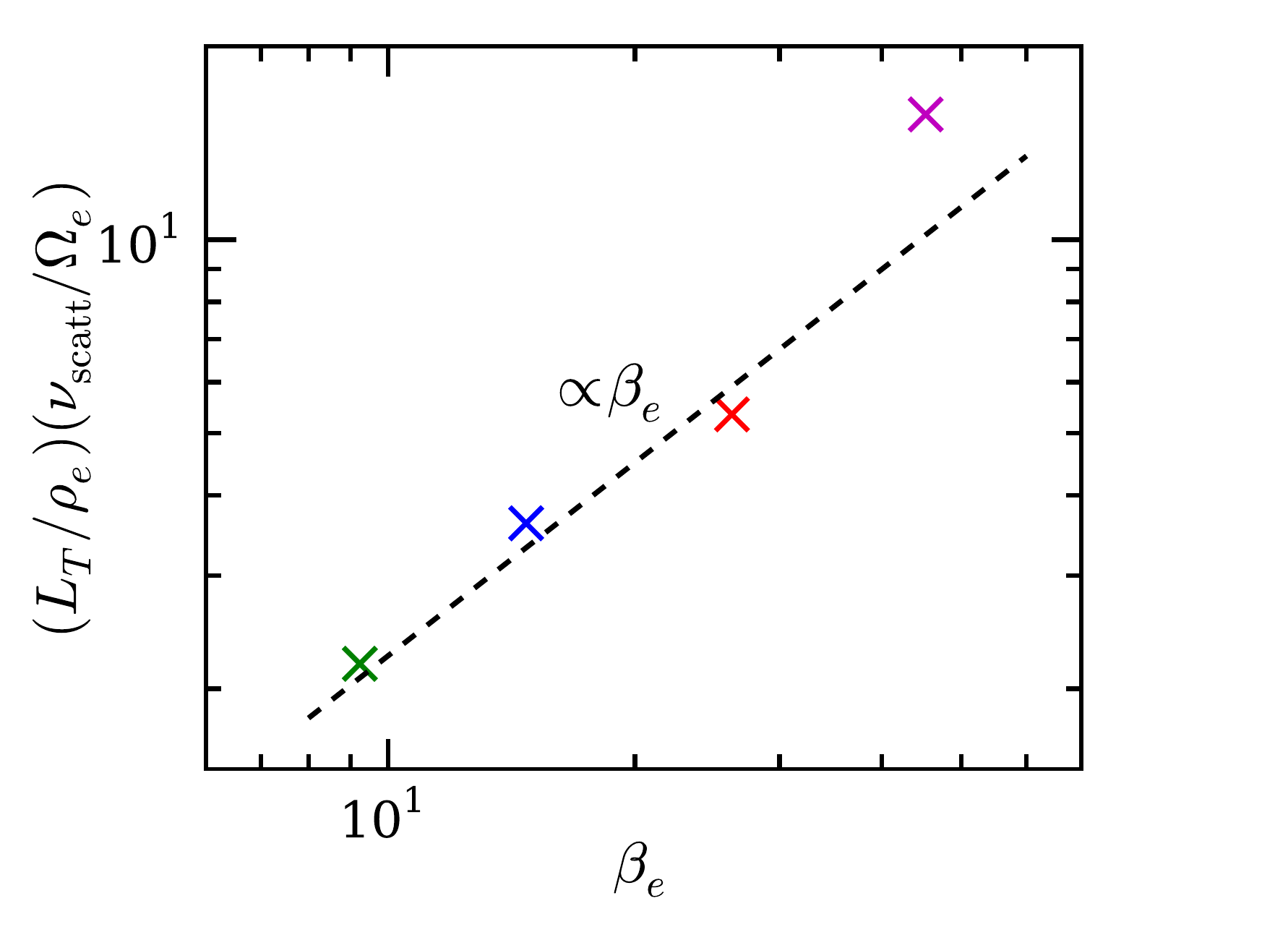}   
    \caption{The electron scattering rate multiplied by $L_T/\rho_e$ (which varies in different runs because of the different final temperature profiles and equals $\sim 100$ on average) as a function of $\beta_e$.  The temperature scale length $L_T=T/\partial_x T$ and the electron Larmor radius $\rho_e$ have been averaged over the simulation domain. We have corrected for the small variation of the effective $\beta_e$ in different runs, as in \figref{fig:simQ}. The dashed line shows the prediction \exref{eq:nuscatt} based on Bohm diffusion combined with the whistler marginality condition.} 
\label{fig:nubeta}
\endminipage\hfill
\minipage[t][][t]{0.45\textwidth}
\centering
 \includegraphics[width=68mm]{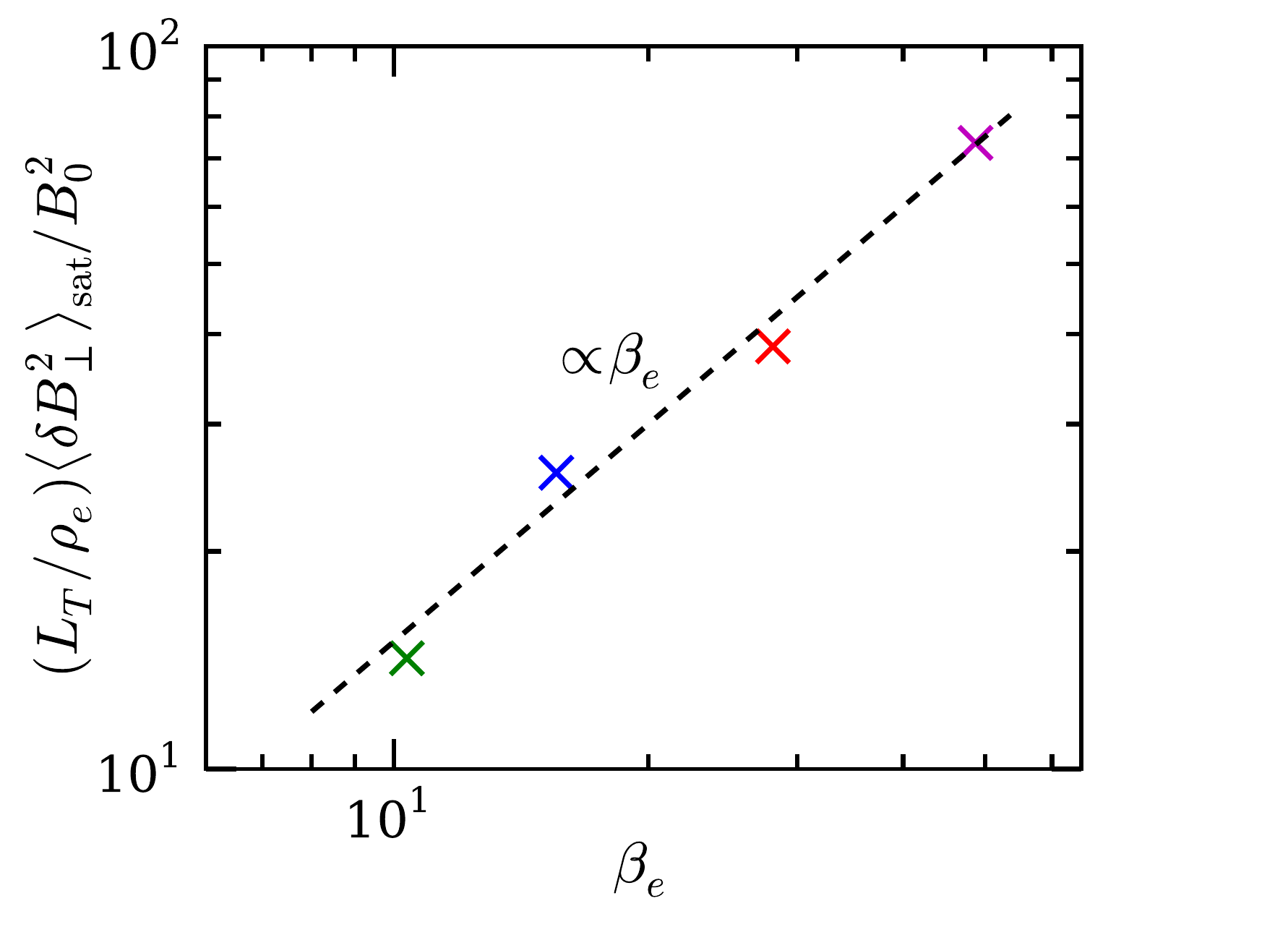}
    \caption{The saturated perpendicular magnetic-field energy density multiplied by $L_T/\rho_e$ (different in different runs, $\sim 100$) as a function of $\beta_e$. The field is averaged over the right third of the computation domain, where it is fully saturated and homogeneous, and also over several time snapshots. The temperature scale length $L_T=T/\partial_x T$ and the electron Larmor radius $\rho_e$ are averaged similarly. We have corrected for the effective $\beta_e$ in the averaging region. The dashed line is a comparison with the prediction \exref{eq:satfield} based on Bohm diffusion combined with the whistler marginality condition.}     
\label{fig:B2beta}
\endminipage
\end{figure}

In \secref{sec:satmf}, based on Bohm diffusion and the whistler marginality condition \exref{eq:epsbeta}, 
we estimated the effective pitch-angle scattering rate as 
a function of $\beta_e$ and the ratio of the electron Larmor radius $\rho_e$ and 
temperature scale length $L_T$. In order to test this prediction, we obtain 
the scattering rate $\nu_{\rm scatt}$ from the simulations by tracing a large number ($\sim 10000$) of test electrons, 
calculating their mean spatial spread along the mean magnetic field, and estimating 
the parallel diffusion coefficient $D_{\parallel} = (1/3) \vth^2 / \nu_{\rm scatt}$ for runs with different $\beta_e$ as follows:
\beq
D_{\parallel} = \frac{1}{2} \frac{d \langle [x_i(t_{\rm diff}) - x_i(0)]^2 \rangle}{dt},
\eeq
where $x_i$ is the $x$ coordinate of particle $i$ and time $t_{\rm diff}$ is taken sufficiently long for diffusion to settle.
We then average 
$L_T=T/\partial_x T$ and $\rho_e$ 
over the computation domain for each run and plot 
$\nu_{\rm scatt} L_T / \rho_e$ as a function of $\beta_e$. The result is shown in 
\figref{fig:nubeta}. There is a good agreement with the qualitative prediction that 
testifies in favor of quasilinear saturation.  Calculating the diffusion coefficient the way we do
for the run with the highest $\beta_e=40$ is likely not fully consistent due to the large variations 
of magnetic perturbations across the box and their large amplitude.    


We can also check if the simple estimate of the saturated magnetic 
field \exref{eq:satfield} based on Bohm diffusion applies to 
the simulated field. 
We average the perpendicular magnetic-energy density over the right third of the 
box, where it is fully saturated and homogeneous in all the runs, and over several 
time snapshots. $L_T$ and $\rho_e$ are again averaged analogously. Similar to 
the scattering rate, we plot the magnetic-energy density multiplied by  $L_T/\rho_e$ 
in  \figref{fig:B2beta}. The Bohm formula describes the simulation results well 
for practically all $\beta_e$. 


\section{Discussion}
\label{sec:disc}

\subsection{Relevance to clusters of galaxies}

\label{sec:clusters}

The intracluster medium (ICM) is a hot rarefied plasma with the typical 
range of $\beta_e$ (recall that the total $\beta_{\rm tot}=\beta_e+\beta_i$ often used in literature 
is approximately twice larger) between several 
tens and several hundreds (see, e.g., \citealt{Kuchar2011}). Thermal conduction 
may or may not play a role in several contexts. One is affecting global radial temperature 
profiles, the knowledge of which is necessary to calculate cluster masses from X-ray 
observations. In this case, 
temperature gradients are small, i.e., temperature scale lengths are long, of the 
order of several hundreds of kpc. Another is the possibility of existence 
of smaller-scale temperature substructure, including cold fronts, where temperature 
gradients can be rather large \citep[e.g.,][]{Ichi2015, Wang2016}, with the temperature scale length approaching 
the classical Spitzer mean free path ($\sim$ 1-10 kpc). 

The suppression of heat flux caused by whistlers is important only if the 
upper limit on the flux imposed by the instability, $q_{{\parallel}{\rm w}}$, turns out 
to be smaller than the Spitzer heat flux \citep{Spitzer1962}. 
The modified Spitzer heat flux $q_{\parallel {\rm S}}$, which includes saturation of 
the flux when large gradients are present on scales smaller than the electron 
Coulomb mean free path, 
can be expressed conveniently as \citep{Cowie1977,Sarazin}
\beq
\label{eq:qs}
q_{{\parallel}{\rm S}} \approx 0.5 m_e n \vth^3 \frac{\lambda_e}{L_T+4\lambda_e},
\eeq          
where $L_T$ is the parallel temperature gradient scale, and $\lambda_e$ is the mean free path for the electron energy exchange. Equation~(\ref{eq:qs}) interpolates between the classical collisional regime $\lambda_e\ll L_T$ and 
the collisionless saturated flux at an infinite temperature gradient. The latter is obtained by assuming a hot plasma adjoining a cold absorber, with a self-consistent electric field set up to stop the current. The free molecular conductivity in this case 
is taken to be reduced by the electric field by the same factor of 0.4 as in the classical case \citep{Spitzer1962}. Other estimates give values of saturated flux by a factor of a few larger (see \citealt{Cowie1977}). Thus, our estimate of suppression compared to the unmagnetized saturated heat flux may be considered conservative.
The ratio of the two fluxes is the suppression factor $\Sw<1$, 
\beq
S_{\rm w}=\frac{q_{{\parallel}{\rm w}}}{q_{{\parallel}{\rm S}}} \approx \frac{3}{\beta_e} \left (\frac{L_T}{\lambda_e} + 4 \right ). 
\eeq
It is easy to write an expression for the effective parallel heat flux $q_{\parallel {\rm eff}}$ that interpolates between the Spitzer flux, when the suppression factors are close to unity, and the flux controlled by whistlers, when suppression is large:
\beq
\label{eq:qinterp}
q_{\parallel {\rm eff}} = \frac{q_{{\parallel}{\rm S}}}{1+\beta_e / 3 (L_T/\lambda_e + 4)^{-1}}\approx \frac{0.5 m_e n \vth^3}{L_T/\lambda_e + \beta_e /3 + 4}.
\eeq
Equation~(\ref{eq:qinterp}) can be used for heat-flux estimates in problems where 
plasma kinetic effects are otherwise ignored, or as a simple subgrid model in numerical simulations.

For the typical parameters of the hot ICM, the energy-exchange electron mean free path is \citep[e.g.,][]{Sarazin}
\beq
\lambda_e \approx 20~{\rm kpc} \left ( \frac{T}{10^8~{\rm K}} \right )^2 \left ( \frac{n}{10^{-3}~{\rm cm}^{-3}} \right )^{-1}.
\eeq
Let us assume $\beta_e \sim 100$ to estimate the maximum suppression of thermal conduction 
possible. The maximum suppression is reached at $L_T \ll \lambda_e$ where $\Sw \sim 1/10$. 
Such small scale lengths of the temperature gradient along the mean magnetic field, however, are unrealistic in galaxy clusters. 
Even in cold fronts, where $L_T\lesssim\lambda_e$, the magnetic field is very likely draped 
perpendicular to the gradient, and thus the parallel gradient scale should be 
appreciably larger. At larger scales, suppression drops linearly: 
$\Sw \sim 1/4$ at 100 kpc, $\Sw \sim 1/2$ at 300 kpc, and no suppression at $\sim$ 600 kpc. 
We conclude, therefore, that in general, the suppression factors caused by the whistler 
instability are rather modest, and are unlikely to affect global radial 
temperature profiles strongly, or, e.g., cut off cool cores from the heat supply from the 
outer hot regions in the absence of other suppression mechanisms. The effect can 
become more important if there are strong thermal gradients on intermediate and small scales $\lesssim$100 kpc, 
where it becomes of the same order as the suppression by the mirror instability 
\citep{Komarov2016} in case these scales are turbulent. The combination of the two effects 
is capable of producing suppression of order $1/30$, which is sufficient to 
explain the variety of substructures typically observed in the ICM. 

Finally, we comment on the possible relevance of the heat-flux suppression by whistlers to the thermal instability in galaxy clusters. Cold ($T\sim10^4$ K) H$\alpha$ filaments  are ubiquitously found in cluster cores \citep[e.g.,][]{McDonald2010}. Such structures are thought to be produced either by dragging the cold material out from central galaxies by buoyant radio bubbles \citep[e.g.,][]{Churazov2001,Fabian2003}, or by runaway cooling in the ICM \cite[e.g.,][]{McCourt2012}.  
In the latter case, during the early phase of a local thermal instability, when the plasma is not yet significantly compressed, 
so $\beta$ is still large, the extra suppression provided by whistlers could promote 
the development of the instability along the magnetic-field lines (see \citealt{Boehringer1989} for a model of spherical cold clump formation in a non-magnetized plasma and \citealt{Sharma2010} for the magnetized case). 
The spatial scale at which cooling is balanced by thermal conduction is known as the Field length $\lambda_F$ \citep{Field1965}. 
Assuming an initial spectrum of density perturbations 
dominated by small scales ($\lesssim 1$ kpc), thermal conduction begins to smear the perturbations along the 
magnetic-field lines on scales below $\lambda_F$, producing filamentary structures (alternatively, the elongated shape 
can be produced by motions in a stratified atmosphere with no magnetic field and no conduction; 
see, e.g., \citealt{McCourt2012}). While later in the evolution the density perturbations can 
continue to fragment into much shorter segments due to the decrease of $\lambda_F$, the large-scale 
coherence (see \citealt{Fabian2008} for a high-resolution observational example) set by the parallel heat fluxes during the early phase should be preserved. 
In the perpendicular direction, there is no conduction, and 
the gas is compressed until it is stabilized by, e.g., the pressure support of cosmic rays or magnetic fields. Therefore, if the length of filaments in the thermal instability scenario is set mainly by parallel conduction, its suppression may lead 
to shorter filaments.

\subsection{Limitations of the model}

\subsubsection{Importance of ions}
\label{sec:ions}

In our simulations and theoretical model, ions do not participate in the instability and form 
a charge-neutralizing background. However, the phase speed of 
the unstable whistlers, 
$v_{\rm ph}\sim \vthe/\beta_e$, approaches the ion thermal speed, $\vthi=(m_e/m_i)^{1/2} \vthe$, 
at $\beta_e \sim 40$. Therefore, oblique whistlers with non-zero parallel electric field may be 
damped by the ions via Landau resonance. 

\begin{figure}
\centering
\includegraphics[width=\textwidth]{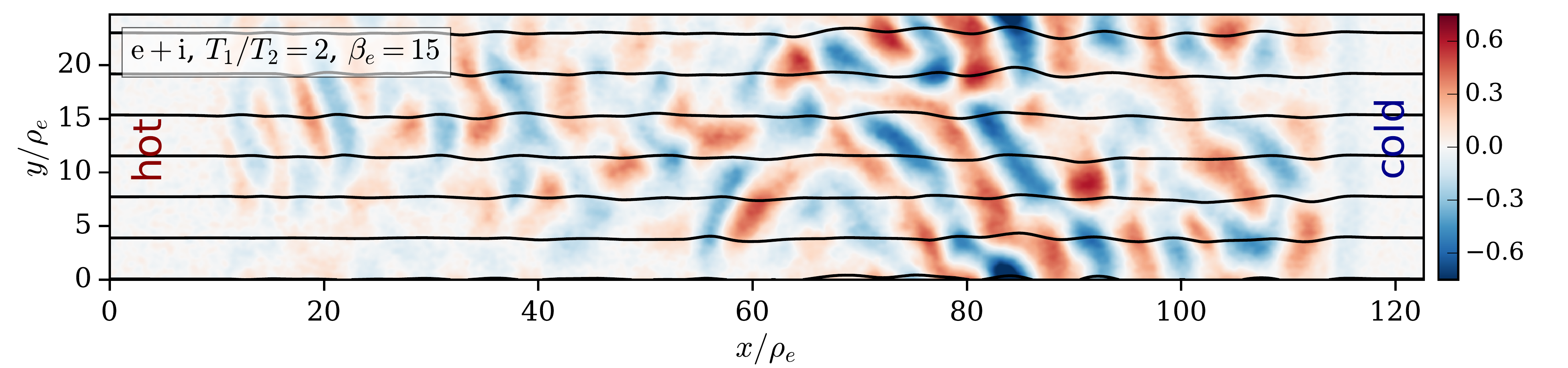}
\caption{The spatial structure of the $z$-component of the magnetic field generated by the heat-flux-induced whistler instability in the run with ions. The mass ratio is $m_i/m_e=225$, the electron plasma 
$\beta_e=15$, so the ions are roughly in Landau resonance with the whistlers. This run 
can be compared with the corresponding electron-only run (second panel from the top in \figref{fig:simB}).}
\label{fig:simBion}
\end{figure} 

In order to test how important the ion physics might be, we have performed a run with 
mass ratio $m_i/m_e=225$, $\beta_e=15$, and $T_1/T_2=2$, i.e., with the ions roughly in resonance 
with the unstable modes. All the other parameters of the simulations have been kept unchanged 
compared with the corresponding main electron-only run. Note that the ion Larmor radius in the new run is rather large, 
$\rho_i\approx 300$ cells. For studying the effect of Landau resonance, however, the ion Larmor scale 
should not be important as long as $\kpd \rho_i \gg 1$.     

The ions do produce a noticeable change in the evolution of the instability. Namely, they 
somewhat delay full saturation. However, the final spatial structure of the magnetic field 
looks similar to the corresponding electron-only run, with the saturated field amplitude only 
$\sim 15$\% smaller (see \figref{fig:simBion}). The heat flux at saturation is roughly 20\% larger. We have not explored 
the full range of $\beta_e$ and $T_1/T_2$ including the ions, but our test run gives us some confidence 
that even close to the Landau resonance, the ions do not seem to choke the instability, 
and our main results remain correct within factors of order unity.

In addition to the above, at sufficiently high $\beta_e \sim m_i/m_e$, the cyclotron resonance with the ions, 
$\omega\sim\Omega_e/\beta_e \sim  \Omega_i$ comes into play, and our theoretical assumptions may break down. But even in this case, simple qualitative arguments based on the analysis of the whistler dispersion relation 
for the case of small propagation angles can be made to show that such damping should not exceed the electron cyclotron damping, and, therefore, the suppression of heat flux should not be 
affected radically. First, it is safe to assume that the real part of the whistler dispersion relation $D(\omega, \vc{k})=0$ is 
largely dominated by the electron contribution (see, e.g., \citealt{Treumann}). Then, we can compare 
the imaginary parts $D_{{\rm im},e}$ and $D_{{\rm im},i}$ (that lead to damping) due to the electrons and ions respectively \citep{Treumann}:
\bea
D_{{\rm im},e} &=& \frac{\pi^{1/2} \omega_{pe}^2}{\kpl v_{{\rm th} e \parallel} \omega} \exp \left [ -\left ( \frac{ \Omega_e-\omega} { \kpl v_{{\rm th} e \parallel} } \right )^2 \right ],\\
D_{{\rm im},i} &=& \frac{\pi^{1/2} \omega_{pi}^2}{\kpl v_{{\rm th} i \parallel} \omega} \exp \left [ -\left ( \frac{ \Omega_i+\omega} { \kpl v_{{\rm th} i \parallel}} \right )^2 \right ].
\eea
Using $(\Omega_i+\omega) /  \kpl v_{{\rm th} i \parallel} \sim \Omega_i /  \kpl v_{{\rm th} i \parallel} \sim (m_e/m_i)^{1/2} \ll 1$ and $(\Omega_e-\omega) /  \kpl v_{{\rm th} e \parallel}  \sim \Omega_e / \kpl v_{{\rm th} e \parallel} \sim 1$, we can obtain the ratio 
\beq
D_{im,e} / D_{im,i} \sim (m_i/m_e)^{1/2} e^{-1} \sim 10. 
\eeq
We see now that the ion cyclotron damping term is significantly smaller than the electron one. 
This can be understood if one notes that in cyclotron resonance $\omega \sim \Omega_i$, an ion 
travels parallel distance $\sim \rho_i \gg \rho_e \sim \kpl^{-1}$ over one gyration period, thus intersecting 
many parallel whistler wavelengths and significantly reducing the efficiency of the wave-ion interaction compared to the electron cyclotron resonance.  
This quick calculation indicates that even at $\beta_e\sim 2000$ our model may still be usable\footnote{As a side note, in the limit of a non-magnetized plasma the whistler instability transforms into the Weibel-like Ramani-Laval instability \citep{RL1978,LE1992}.}.

We should stress that the above discussion does not deny the need for proper numerical modeling of the heat-flux-induced whistler instability with ion physics, but is aimed to support at least the qualitative usefulness of our model in light of the seemingly problematic issue of ion resonances appearing already at modest $\beta_e\lesssim100$. 

\subsubsection{Reaching regime of small magnetic perturbations}

In astrophysical environments, 
temperature-gradient scales exceed electron Larmor radii by many orders of magnitude. Therefore, 
the heat-flux-induced whistler instability in these environments saturates at a very low level, as described by \exref{eq:satfield}. Numerically, 
such a regime is demanding to simulate. First, if one needs to study suppression of thermal conduction 
by the instability, $\beta_e$ should be sufficiently large, at least $\beta_e > 10$, as we have shown first in 
\secref{sec:heatflux}, and then, in relation to galaxy clusters, in \secref{sec:clusters}. Then either the 
temperature difference between the hot and cold walls should be set small, or the computation domain 
should be made long to minimize $\rho_e/L_T$. The former is problematic because the initial collisionless 
Knudsen heat flux in the absence of electromagnetic fields can be already too small, below the marginal 
level of the instability. Thus, we are left with the need to use a rather long simulation box. This is also dictated 
by the requirement that wave advection should not affect the saturation level in most of the box, as we have discussed 
in \secref{sec:fieldstruct}. Our simulations show, however, that even the rather long box that we use provides 
only a narrow range of $\beta_e$ in which magnetic perturbations could be assumed small.  
Already at $\beta_e=25$ perturbations reach amplitudes $\delta B / B_0 \sim 1$. In this regard, we need to 
be confident that the suppression mechanism in the simulation has been identified correctly. 

\subsubsection{Possibility of alternative saturation mechanisms}

Different nonlinear saturation mechanisms can by ruled out to a certain extent by comparison of 
nonlinear wave damping rates with the quasilinear damping rate \exref{eq:damp}. Additionally, one can compare 
the behavior of the predicted nonlinear saturation levels as functions of $\beta_e$ by equating the linear growth rate and nonlinear damping rate. 
Such predictions are usually only possible under a number of drastically simplifying assumptions.  

Let us consider nonlinear mode coupling first. \citealt{LE1992} proposed it as a saturation mechanism 
of the instability. \citealt{Levinson1992} calculated the rate of nonlinear whistler mode coupling by using a perturbation approach in the approximation of near-parallel, $\kpd/\kpl \ll 1$, propagation (which is a questionable assumption at best in our case, but also the one that permits obtaining a qualitative estimate):
\beq
\label{eq:MC}
\gamma_{\rm MC} \sim \frac{\nu_{\rm scatt}}{\beta_e} \sim \frac{\vth}{\epsilon \beta_e L_T},
\eeq
where we have used $\nu_{\rm scatt} = \vth / \mfp = \vth / \epsilon L_T$ and $\epsilon = \mfp / L_T$. By requiring the linear growth rate $\gamma_{\rm lin} \sim \epsilon \Omega_e$ to be balanced by nonlinear damping, we get 
\beq
\label{eq:epsMC}
\epsilon = (\rho_e / \beta_e L_T)^{1/2},
\eeq 
and 
\beq
\gamma_{\rm MC} \sim \frac{\vth}{(\beta_e \rho_e L_T)^{1/2}}.
\eeq
The quasilinear damping rate is $\gamma_{\rm QL} \sim \beta_e^{-1} \Omega_e$, and is seen to be a factor of 
\beq
\frac{\gamma_{\rm QL}}{\gamma_{\rm MC}} \sim \left ( \frac{L_T}{\beta_e \rho_e}  \right )^{1/2}
\eeq
larger than the rate of mode coupling. This factor is very large in astrophysical plasmas. In the simulation, however, it approaches unity at higher $\beta_e$. In order to make a more detailed comparison, we can estimate the saturation level of magnetic perturbations from Bohm diffusion, $\delta B^2 / B_0^2\sim \nu_{\rm scatt} / \Omega_e$, and substitute $\nu_{\rm scatt}$ using equations~(\ref{eq:MC}) and (\ref{eq:epsMC}):
\beq
\frac{\delta B^2}{B_0^2} \sim \left ( \frac{\beta_e \rho_e}{L_T}  \right )^{1/2}.
\eeq
We see that the saturation level scales differently with $\beta_e$ and $\rho_e/L_T$ than in our simulation. This gives us some confidence that the saturation is not regulated by mode coupling in our results.  The nonlinear Landau damping rate can also be estimated in the same approximation, but is found to be an order of $\beta_e^2$ smaller than the mode coupling term \citep{Levinson1992}.

Alternatively, when magnetic perturbations grow large and electron orbits become distorted, the instability may saturate by resonance broadening. Resonance broadening leads to leakage of particles out of the resonance until saturation is reached.  This is modeled by the delta function in \exref{eq:gammak} substituted by the Lorenz function with a width set by electron scattering off magnetic perturbations. Using frequencies in place of momenta,
\beq
\delta (\omega - \kpl v_{\parallel} \pm \Omega_e) \longrightarrow \frac{\delta_{\rm r}}{(\omega-\kpl v_{\parallel} \pm \Omega_e)^2 + \delta_{\rm r}^2 },
\eeq   
where $\delta_{\rm r}=(\kpl^2 D/3)^{1/3}$ is the frequency half-width of the particle resonance and $D$ is the electron velocity diffusion coefficient (see, e.g., \citealt{Treumann}). The instability is stabilized when the resonance broadens beyond the width of the region of resonant particles in velocity space $\Delta \vpl$, which can be roughly estimated as $\Delta \vpl \sim \vth$ (recall that the spectrum of excited waves is wide, $\Delta \kpl/\kpl \sim 1$, and so is the resonant region in velocity space). Taking $D\sim\vth^2 \nu_{\rm scatt} \sim  \vth^2 \Omega_e \delta B^2/B_0^2$ from Bohm diffusion (assume that only the direction of electron velocity changes during scattering) as before, we are able to obtain the saturation level from the condition $\delta_{\rm r} \sim \kpl \vth$:
\beq
\frac{\delta B^2}{B_0^2} \sim \frac{\kpl \vth}{\Omega_e} \sim 1,
\eeq
where we have used the fact that the resonant scale is virtually independent of the temperature gradient and $\beta_e$.
Thus, resonant broadening should lead to a saturated magnetic field whose amplitude does not depend on $\beta_e$, in contradiction with our numerical results.

The above qualitative arguments appear to strengthen our point that the main saturation mechanism of the heat-flux-induced whistler instability both in astrophysical environments and our simulations has been identified correctly.

\section{Conclusions}
\label{sec:concl}

Aided by numerical simulations, we have demonstrated that, in the presence of a 
temperature gradient, a weakly collisional high-$\beta$ plasma is susceptible to the whistler 
instability. The instability quickly develops a spectrum of oblique modes that 
are able to scatter the heat-flux-carrying electrons. We have also confirmed 
the quasilinear result
that at saturation, the marginal level of the heat flux is set by the inverse 
plasma $\beta$ rather than by the imposed temperature gradient. The 
numerical results have been shown to be in agreement with simple quasilinear 
arguments, such as the linear scaling of the pitch-angle scattering rate and the saturation level of magnetic perturbations with $\beta_e$. In the 
context of galaxy clusters, the instability can introduce moderate suppression 
factors of thermal conduction $\sim 1/4$ on scales $\sim 100$ kpc if significant variations of temperature 
are found there. We have given a simple expression \exref{eq:qinterp} for the amount of 
suppression of the heat flux as a function of the temperature gradient scale length and 
the plasma beta. This expression can be applied to models in which kinetic effects 
are difficult to implement directly. Combined with the suppression 
by the mirror instability in a turbulent high-$\beta$ plasma, the two effects add up 
causing large suppression factors (several tens in the case of galaxy clusters). 

\section*{Acknowledgements}

SK and EC acknowledge support by grant No. 14-22-00271 from the Russian Science Foundation.
The work of AAS was supported in part by grants from UK STFC and EPSRC. He wishes to thank A. Bott and S. Cowley for very useful conversations.
\hspace{30mm}


\bibliographystyle{jpp}
\bibliography{bibliography1}

\end{document}